\documentclass[aps,superscriptaddress,twocolumn,pra,showpacs]{revtex4-1}
\usepackage{graphicx}
\usepackage{amsmath}
\usepackage{ulem}
\usepackage{hyperref}

\begin{document}

\title{The dynamical Casimir effect in superconducting microwave circuits}

\author{J.R.~Johansson}
\affiliation{Advanced Science Institute, RIKEN, Wako-shi, Saitama 351-0198, Japan}
\author{G.~Johansson}
\affiliation{Microtechnology and Nanoscience, MC2, Chalmers University of
Technology, SE-412 96 G{\"o}teborg, Sweden}
\author{C.M.~Wilson}
\affiliation{Microtechnology and Nanoscience, MC2, Chalmers University of
Technology, SE-412 96 G{\"o}teborg, Sweden}
\author{Franco Nori} 
\affiliation{Advanced Science Institute, RIKEN, Wako-shi, Saitama 351-0198, Japan}
\affiliation{Physics Department, The University of Michigan, Ann Arbor, Michigan 48109-1040, USA}

\date{\today}

\begin{abstract}
We theoretically investigate the dynamical Casimir effect in electrical circuits based on
superconducting microfabricated waveguides with tunable boundary conditions.
We propose to implement a rapid modulation of the boundary conditions 
by tuning the applied magnetic flux through superconducting quantum interference devices (SQUIDs) that are embedded in the waveguide circuits.
We consider two circuits: 
($i$) An open waveguide circuit that corre\-sponds to a single mirror in free space,
and ($ii$) a resonator coupled to a microfabricated waveguide, which corre\-sponds to a single-sided cavity in free space.
We analyze the properties of the dynamical Casimir effect in these two setups
by calculating the generated photon-flux density, output-field correlation functions,
and the quadrature squeezing spectra.
We show that these properties of the output field exhibit signatures unique to the radiation due to the dynamical Casimir effect,
and could therefore be used for distinguishing the dynamical Casimir effect from other types of radiation in these circuits.
We also discuss the similarities and differences 
between the dynamical Casimir effect, in the resonator setup, and downconversion of pump photons in parametric oscillators. 
\end{abstract}


\maketitle

%
\section{Introduction}

Quantum field theory predicts that photons can be created from vacuum fluctuations when the boundary conditions of the field are time-dependent.
This effect, often called the dynamical Casimir effect, was predicted by G.T.~Moore \cite{Moore1970} in 1970, 
in the context of a cavity comprised of two moving ideal mirrors.
In 1976, S.A.~Fulling {\it et al.} \cite{Fulling1976}
showed that a single mirror in free space also generates radiation, when subjected to a nonuniform acceleration.
The role of the moving mirrors in these studies is to impose time-dependent boundary conditions on the electromagnetic fields.
The interaction between the time-dependent boundary condition
and the zero-point vacuum fluctuations 
can result in photon creation,
for a sufficiently strong time-dependence \cite{Barton1993,Dodonov2001,Dodonov2010,Dalvit2010}.

However, it has proven to be a difficult task to experimentally observe the dynamical Casimir effect.
The problem lies in the difficulty in changing the boundary conditions,
e.g., by moving physical objects, such as massive mirrors,
sufficiently fast to generate a significant number of photons.
Although there are proposals (see, e.g., Ref.~\cite{Onofrio2006}) for experimentally observing the dynamical Casimir effect using massive mirrors,
no experimental verification of the dynamical Casimir effect has been reported to date \cite{Dodonov2010}.
In order to circumvent this difficulty
a number of theoretical proposals have suggested
to use experimental setups where the boundary conditions are modulated by some effective motion instead.
Examples of such proposals include to use lasers to modulate the reflectivity of thin semiconductor films \cite{Crocce2004,Braggio2005}
or to modulate the resonance frequency of a superconducting stripline resonator \cite{Segev2007}, 
to use a SQUID to modulate the boundary condition of a superconducting waveguide \cite{Johansson2009}, 
and to use laser pulses to rapidly modulate the vacuum Rabi frequency in cavity QED systems \cite{Gunter2009,DeLiberato2009}.

In this paper we investigate manifestations of the dynamical Casimir effect in superconducting electrical circuits based on microfabricated (including coplanar) waveguides. 
Recent theoretical and experimental developments in the field of superconducting electronics,
which to a large extent is driven by research on quantum information processing \cite{You2005,Wendin2006,Clarke2008},
include the realization of strong coupling between artificial-atoms and oscillators \cite{Chiorescu2004,Wallraff2004,Schoelkopf2008} (so called circuit QED), studies of the ultra-strong coupling regime in circuit QED \cite{Ashhab2010}, 
single-artificial-atom lasing \cite{Astafiev2007,Ashhab2008},
Fock-state generation and state tomography \cite{Hofheinz2008, Liu2004}.
Also, there has recently been an increased activity in studies of multimode quantum fields in superconducting circuits,
both theoretically and experimentally, see e.g., Refs.~\cite{Zhou2008,Abdumalikov2008,Astafiev2010,Liao2010},
and in experimental work on frequency-tunable resonators \cite{Palacois2008,Yamamoto2008,Lehnert2008,Sandberg2008}.
These studies exemplify how quantum-optics-like systems can be implemented in superconducting electrical circuits \cite{Buluta2009},
where waveguides and resonators play the roles of light beams and cavities, 
and Josephson-junction based artificial atoms play the role of natural atoms in the original quantum optics setups.

\begin{figure*}[t]
\includegraphics[width=14.5cm]{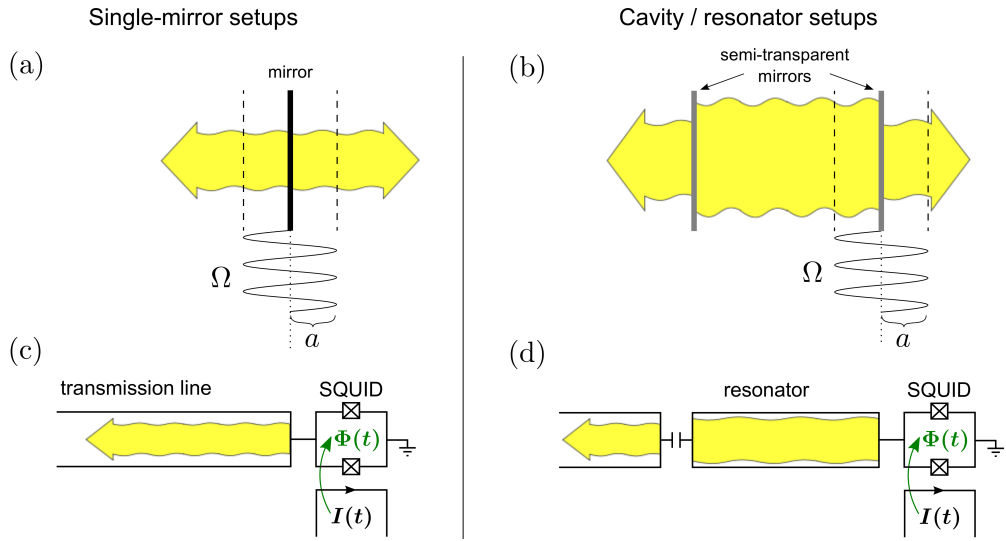}
\caption{(color online)
Schematic illustration of the dynamical Casimir effect, in the case of a single oscillating mirror in free space (a),
and in the case of a cavity in free space, where the position of one of the mirrors oscillates (b). 
In both cases, photons are generated due to the interplay between the time-dependent boundary conditions imposed by the moving mirrors, and the vacuum fluctuations. Here, $\Omega$ is the frequency of the oscillatory motion of the mirrors, and $a$ is the amplitude of oscillations.
The dynamical Casimir effect can also be studied in electrical circuits.
Two possible circuit setups that correspond to the quantum-optics setups (a) and (b) are shown schematically in (c) and (d), respectively.
In these circuits, the time-dependent boundary condition imposed by the SQUID corresponds to the motion of the mirrors in (a) and (b).
}
\label{fig:schematic-dce}
\end{figure*}

Here, we theoretically investigate the possibility to exploit these recent advances
to realize a system \cite{Johansson2009} where the dynamical Casimir effect can be observed experimentally in an electrical circuit. 
We consider two circuit configurations, see Fig.~\ref{fig:schematic-dce}(c)-(d),
for which we study the dynamical Casimir effect in the broadband and narrowband limits, respectively.
We analyze the properties of the radiation due to the dynamical Casimir effect in these systems,
and we identify a number of signatures in experimentally measurable signals
that could be used to distinguish the radiation due to the dynamical Casimir effect from other types of radiation,
such as thermal noise.

The dynamical Casimir effect has also previously been discussed in the context of superconducting electrical circuits in 
Ref.~\cite{Takashima2008}.
Another related theoretical proposal to use superconducting electrical circuits
to investigate photon creation due to nonadiabatic changes in the field parameters was presented in \cite{Nation2009}, 
where a circuit for simulating the Hawking radiation was proposed. 
In contrast to these works, here we exploit the demonstrated fast tunability of the boundary conditions for a one-dimensional
electromagnetic field, achieved by terminating a microfabricated waveguide with a SQUID \cite{Sandberg2008}.
(See, e.g., Ref.~\cite{Crispino2008} for a review of the connections between the dynamical Casimir effect,
the Unruh effect and the Hawking effect).

%
%
This paper is organized as follows:
In Sec.~\ref{sec:review}, we briefly review the dynamical Casimir effect. 
In Sec.~\ref{sec:sc-dce-single-mirror}, we propose and analyze an electrical circuit, Fig.~\ref{fig:schematic-dce}(c), based on an open microfabricated waveguide, for realizing the dynamical Casimir effect, and we derive the resulting output field state.
In Sec.~\ref{sec:sc-dce-resonator}, we propose and analyze an alternative circuit, Fig.~\ref{fig:schematic-dce}(d), featuring a waveguide resonator.
In Sec.~\ref{sec:measurement-setups}, we investigate various measurement setups that are realizable in electrical circuits in the microwave regime, and we explicitly evaluate the photon-flux intensities and output-field correlation functions for the two setups introduced in Sec.~III and IV.
In Sec.~\ref{sec:parametric-oscillator}, we explore the similarities between the dynamical Casimir effect,
in the resonator setup,
and the closely related parametric oscillator with a Kerr-nonlinearity.
Finally, a summary is given in Sec.~\ref{sec:summary}.

%
\section{Brief review of the dynamical Casimir effect}
\label{sec:review}

\begin{table*}[tbp]
\label{tbl:static-vs-dynamical-casimir}
\begin{center}
\begin{tabular}{c|c|c}
\hline\hline
& Static Casimir Effect & Dynamical Casimir Effect  \\ \hline\hline
Description & 
\parbox{7.5cm}{\vspace{0.2cm}Attractive force between two conductive plates in vacuum.\vspace{0.2cm}} & 
\parbox{7.5cm}{\vspace{0.2cm}Photon production due to a fast modulation of boundary conditions.\vspace{0.2cm}}
\\ \hline
Theory & 
\parbox{7.8cm}{
\vspace{0.2cm}
Casimir (1948) \cite{Casimir1948}, 
Lifshitz (1956) \cite{Lifshitz1956}.
\vspace{0.2cm}} & 
\parbox{7.5cm}{\vspace{0.2cm}Moore (1970) \cite{Moore1970}, Fulling {\it et al.} (1976) \cite{Fulling1976}.\vspace{0.2cm}}
\\ \hline
Experiment & 
\parbox{7.8cm}{\vspace{0.2cm}
Sparnaay (1958) \cite{Sparnaay1958}, 
van Blokland {\it et al.}. (1978) \cite{vanBlokland1978}, 
Lamoreaux (1997) \cite{Lamoreaux1997}, 
Mohideen {\it et al.}. (1998) \cite{Mohideen1998}.
\vspace{0.2cm}} & 
---
\\ \hline\hline
\end{tabular}
\end{center}
\caption{Brief summary of early work on the static and the dynamical Casimir effect. The static Casimir effect has been experimentally verified, but experimental verification of the dynamical Casimir effect has not yet been reported \cite{Dodonov2010}.}
\end{table*}

\subsection{Static Casimir effect}

Two parallel perfectly conducting uncharged plates (ideal mirrors) in vacuum attract each other with a force known as the Casimir force. 
This is the famous static Casimir effect, predicted by H.B.G.~Casimir in 1948 \cite{Casimir1948}, 
and it can be interpreted as originating from vacuum fluctuations
and due to the fact that the electromagnetic mode density is different inside and outside of the cavity formed by the two mirrors.
The difference in the mode density results in a radiation pressure on the mirrors,
due to vacuum fluctuations,
that is larger from the outside than from the inside of the cavity,
thus producing a force that pushes the two mirrors towards each other.
The Casimir force has been thoroughly investigated theoretically, 
including different geometries, 
nonideal mirrors, 
finite temperature, 
and it has been demonstrated experimentally in a number of different situations
(see, e.g., Refs.~\cite{Lamoreaux1997,Mohideen1998,Bressi2002}). 
For reviews of the static Casimir effect, see, e.g., Refs.~\cite{Milonni1994,Lamoreaux1999,Capasso2007}.

\subsection{Dynamical Casimir effect}

The dynamical counterpart to the static Casimir effect occurs when one or two of the mirrors move.
The motion of a mirror can create electromagnetic excitations, 
which results in a reactive damping force that opposes the motion of the mirror
\cite{Jaekel1992}.
This prediction can be counter-intuitive at first sight, 
because it involves the generation of photons ``from nothing'' (vacuum) with uncharged conducting plates,
and it has no classical analogue.
However, in the quantum mechanical description of the electromagnetic field,
even the vacuum contains fluctuations,
and the interaction between these fluctuations and the time-dependent boundary conditions can create excitations (photons)
in the electromagnetic field.
In this process,
energy is taken from the driving of the boundary conditions to excite vacuum fluctuations to pairs of real photons,
which propagate away from the mirror.

The electromagnetic field in a one-dimensional cavity with variable length
was first investigated quantum mechanically by G.T.~Moore \cite{Moore1970}, in 1970.
In that seminal paper,
the exact solution for the electromagnetic field in a one-dimensional cavity with an arbitrary cavity-wall trajectory
was given in terms of the solution to a functional equation, known as Moore's equation.
Explicit solutions to this equation for specific mirror motions has been the topic of numerous subsequent papers,
including perturbative approaches valid in the short-time limit \cite{Dodonov1990},
asymptotic solutions for the long-time limit \cite{Dodonov1993}, 
an exact solution for a nearly-harmonically oscillating mirror \cite{Law1994}, 
numerical approaches \cite{Cole1995}, 
and renormalization group calculations valid in both the short-time and long-time limits \cite{Dalvit1998,Dalvit1999}.
Effective Hamiltonian formulations were reported in \cite{Razavy1989,Law1993,Law1995},
and the interaction between the cavity field and a detector was studied in \cite{Dodonov1995,Dodonov1996}.
The dynamical Casimir effect was also investigated in three-dimensional cavities \cite{Dodonov1995,Dodonov1996,Crocce2001},
and for different types of boundary conditions \cite{Alves2006,Alves2010}. 
The rate of build-up of photons depends in general on the exact trajectory of the mirror,
and it is also different in the one-dimensional and the three-dimensional case.
For resonant conditions, i.e., where the mirror oscillates with twice the natural frequency of the cavity,
the number of photons in a perfect cavity grows exponentially with time \cite{Meplan1995}.

An alternative approach that focuses on the radiation generated by a nonstationary mirror,
rather than the build-up of photons in a perfect cavity,
was developed by S.A.~Fulling {\it et al.}~\cite{Fulling1976}, in 1976.
In that paper, it was shown that a single mirror in one-dimensional free space (vacuum) subjected to a nonuniform acceleration also produces radiation. 
The two cases of oscillatory motion of a single mirror, and a cavity with walls that oscillate in a synchronized manner,
were studied in Refs.~\cite{Lambrecht1996,Lambrecht1998}, 
using scattering analysis.
The radiation from a single oscillating mirror was also analyzed in three dimensions~\cite{Neto1996}.

Table I briefly compares the static and the dynamical Casimir effect.
See, e.g., Refs.~\cite{Barton1993,Dodonov2001,Dodonov2010,Dalvit2010} for extensive reviews of the dynamical Casimir effect.

\subsection{Photon production rate}
The rate of photon production of an oscillating ideal mirror in one-dimensional free space \cite{Lambrecht1996},
see Fig.~\ref{fig:schematic-dce}(a), is, to first order,
\begin{equation}
\label{eq:review_N}
 \frac{N}{T} = \frac{\Omega}{3\pi}\left(\frac{v}{c}\right)^2,
\end{equation}
where $N$ is the number of photons generated during the time $T$,
$\Omega$ is the oscillation frequency of the mirror,
$v = a\Omega$ is the maximum speed of the mirror,
and $a$ is the amplitude of the mirror's oscillatory motion.
From this expression it is apparent that to achieve significant photon production rates, the ratio $v/c$ must not be too small (see, e.g., Table II).
The maximum speed of the mirror must therefore approach the speed of light.
The spectrum of the photons generated in this process has a distinct parabolic shape, between zero frequency and the driving frequency $\Omega$,
\begin{equation}
\label{eq:review_n_w}
 n(\omega) \propto \left(\frac{a}{c}\right)^2\omega(\Omega-\omega).
\end{equation}
This spectral shape is a consequence of the density of states of electromagnetic modes in one-dimensional space, and the fact that photons are generated in pairs with frequencies that add up to the oscillation frequency of the boundary: $\omega_1 + \omega_2 = \Omega$.

\begin{table}[tbp]
\label{tbl:photon-production-rates}
\begin{center}
\begin{tabular}{c|c|c|c}
\hline\hline
Setup & 
\parbox{1.75cm}{\vspace{0.2cm} Amplitude \\ $a$ (m) \vspace{0.1cm}} &
\parbox{1.75cm}{\vspace{0.2cm} Frequency \\ $\Omega$ (Hz) \vspace{0.1cm}} & 
\parbox{1.75cm}{\vspace{0.2cm} Photons \\ $n$ (s$^{-1}$) \vspace{0.1cm}} 
\\ \hline\hline
\parbox{2.0cm}{\vspace{0.2cm} Mirror moved by hand \vspace{0.1cm}} & 
\parbox{1.75cm}{\vspace{0.2cm} 1          \vspace{0.1cm}} & 
\parbox{1.75cm}{\vspace{0.2cm} 1          \vspace{0.1cm}} & 
\parbox{1.75cm}{\vspace{0.2cm} $\sim10^{-18}$ \vspace{0.1cm}}
\\ \hline
\parbox{2.0cm}{\vspace{0.2cm} Mirror on a nano\-mechanical oscillator\vspace{0.1cm}} & 
\parbox{1.75cm}{\vspace{0.2cm} $10^{-9}$  \vspace{0.1cm}} & 
\parbox{1.75cm}{\vspace{0.2cm} $10^{9}$   \vspace{0.1cm}} & 
\parbox{1.75cm}{\vspace{0.2cm} $\sim10^{-9}$  \vspace{0.1cm}}
\\ \hline
\parbox{2.0cm}{\vspace{0.2cm} SQUID-terminated CPW \cite{Johansson2009} \vspace{0.1cm}} & 
\parbox{1.75cm}{\vspace{0.2cm} $10^{-4}$  \vspace{0.1cm}} & 
\parbox{1.75cm}{\vspace{0.2cm} $10^{10}$  \vspace{0.1cm}} & 
\parbox{1.75cm}{\vspace{0.2cm} $\sim10^{5}$   \vspace{0.1cm}}
\\ \hline\hline
\end{tabular}
\end{center}
\caption{The photon production rates, $n=N/T$, for a few examples of single-mirror systems. The order of magnitudes of the photon production rates are calculated using Eq.~(\ref{eq:review_N}). The table illustrates how small the photon production rates are unless both the amplitude and the frequency are large, so that the maximum speed of the mirror $v_{\rm max} = a \Omega$ approaches the speed of light. The main advantage of the coplanar waveguide (CPW) setup is that the amplitude of the effective motion can be made much larger than for setups with massive mirrors that oscillate with a comparable frequency.}
\end{table}

By introducing a second mirror in the setup, so that a cavity is formed, see Fig.~\ref{fig:schematic-dce}(b),
the dynamical Casimir radiation can be resonantly enhanced. 
The photon production rate for the case when the two cavity walls oscillate in a synchronized manner \cite{Lambrecht1996,Lambrecht1998}, is
\begin{eqnarray}
 \frac{N}{T} = Q \frac{\Omega}{3\pi}\left(\frac{v}{c}\right)^2,
\end{eqnarray}
where $Q$ is the quality factor of the cavity.

In the following sections we consider implementations of one-dimensional single- and two-mirror setups
using superconducting electrical circuits. See Fig.~\ref{fig:schematic-dce}(c) and (d), respectively.
The single-mirror case is studied in the context of a semi-infinite waveguide in Sec.~\ref{sec:sc-dce-single-mirror},
and the two-mirror case is studied in the context of a resonator coupled to a waveguide in Sec.~\ref{sec:sc-dce-resonator}.
In the following we consider circuits with coplanar waveguides,
but the results also apply to circuits based on other types of microfabricated waveguides.

%
%
\section{The dynamical Casimir effect in a semi-infinite coplanar waveguide}
\label{sec:sc-dce-single-mirror}

In a recent paper \cite{Johansson2009}, we proposed a semi-infinite superconducting coplanar waveguide terminated by a superconducting interference device (SQUID) as a possible device for observing the dynamical Casimir effect. See Fig.~\ref{fig:cpw-squid-schematic}. 
The coplanar waveguide contains a semi-infinite one-dimensional electromagnetic field,
and the SQUID provides a means of tuning its boundary condition.
Here we present a detailed analysis of this system based on quantum network theory \cite{Yurke1984,Devoret1995}.
We extend our previous work by investigating field correlations and the noise-power spectra of the generated dynamical Casimir radiation, 
and we also discuss possible measurement setups.

%
%
\subsection{Quantum network analysis of the SQUID-terminated coplanar waveguide}

\begin{figure}[t]
\includegraphics[width=8.0cm]{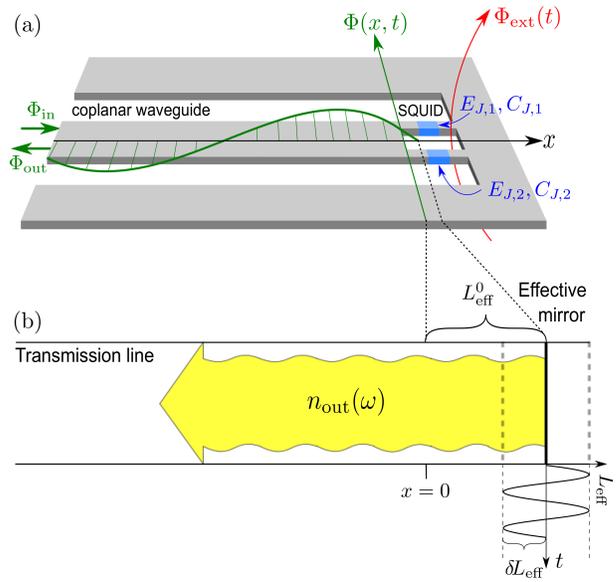}
\caption{(color online) (a) Schematic illustration of a coplanar waveguide terminated by a SQUID. The SQUID imposes a boundary condition in the coplanar waveguide that can be parametrically tuned by changing the externally applied magnetic flux through the SQUID. (b) The setup in (a) is equivalent to a waveguide with tunable length, or to a mirror with tunable position.}
\label{fig:cpw-squid-schematic}
\end{figure}

\begin{figure}[t]
\includegraphics[width=8.0cm]{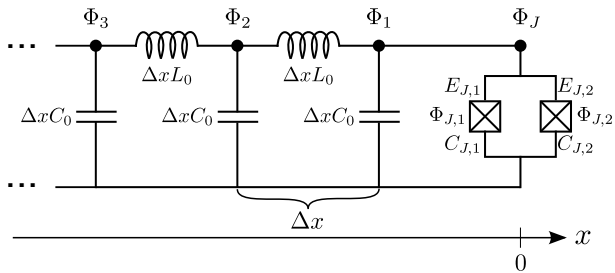}
\caption{Equivalent circuit diagram for a coplanar waveguide terminated by a SQUID. The coplanar waveguide has a characteristic inductance $L_0$ and capacitance $C_0$, per unit length, and it is assumed that it does not have any intrinsic dissipation. The circuit is characterized by the dynamical fluxes $\Phi_i$ and $\Phi_{J,j}$.}
\label{fig:cpw-squid-circuit-diagram}
\end{figure}

In this section we present a circuit model for the proposed device
and we derive the boundary condition of the coplanar waveguide that is imposed by the SQUID (see Fig.~\ref{fig:cpw-squid-schematic}).
The resulting boundary condition is then used in the input-output formalism to solve for the output field state,
in the two cases of static and harmonic driving of the SQUID.
The circuit diagram for the device under consideration is shown in Fig.~\ref{fig:cpw-squid-circuit-diagram},
and the corresponding circuit Lagrangian is 
\begin{eqnarray}
\mathcal{L}
&=&
\frac{1}{2} \sum_{i=1}^{\infty}
\left(
	\Delta x\,C_0(\dot\Phi_{i})^2-\frac{(\Phi_{i+1}-\Phi_{i})^2}{\Delta xL_0}
\right)
\nonumber\\
&+&
\sum_{j=1,2}\left(\frac{C_{J,j}}{2}(\dot\Phi_{J,j})^2 + E_{J,j}\cos\left(2\pi\frac{\Phi_{J,j}}{\Phi_0}\right)\right),
\end{eqnarray}
where $L_0$ and $C_0$ are, respectively, the characteristic inductance and capacitance of the coplanar waveguide (per unit length), and $C_{J,j}$ and $E_{J,j}$ are the capacitance and Josephson energy of the $j$th junction in the SQUID loop.
Here, $\Phi_\alpha$ is the node flux, which is related to the phase $\varphi_\alpha$, at the node $\alpha$, as $\Phi_\alpha = (\Phi_0/2\pi)\varphi_\alpha$, where $\Phi_0 = h/2e$ is the magnetic flux quantum.

%
%
We have assumed that the geometric size of the SQUID loop is small enough such that the SQUID's self-inductance, $L_s$, is negligible compared to the kinetic inductance associated with the Josephson junctions $(\Phi_0/2\pi)^2/E_{J,j}$ (i.e., a term of the form $L_s I_s^2$ has been dropped from the Lagrangian above, where $I_s$ is the circulating current in the SQUID). Under these conditions, the fluxes of the Josephson junctions are related to the externally applied magnetic flux through the SQUID, $\Phi_{\rm ext}$, according to $\Phi_{J,1}-\Phi_{J,2}=\Phi_{\rm ext}$. We can therefore reduce the number of fluxes
used to describe the SQUID
by introducing $\Phi_J=(\Phi_{J,1}+\Phi_{J,2})/2$, and the SQUID effectively behaves as a single Josephson junction \cite{Likharev1986}.

%
%
Under the additional assumption that the SQUID is symmetric, i.e., $C_{J,1}=C_{J,2}=C_J/2$ and $E_{J,1}=E_{J,2}=E_{J}$, the Lagrangian now takes the form
\begin{eqnarray}
\mathcal{L}
&=&
\frac{1}{2} \sum_{i=1}^{\infty} \left( \Delta x\,C_0(\dot\Phi_{i})^2-\frac{(\Phi_{i+1}-\Phi_{i})^2}{\Delta xL_0} \right) 
\nonumber\\
&+&
\frac{1}{2}C_{J}\,(\dot\Phi_J)^2 + E_{J}(\Phi_{\rm ext})\cos\left(2\pi\frac{\Phi_J}{\Phi_0}\right),
\end{eqnarray}
with effective junction capacitance $C_J$ and tunable Josephson energy
\begin{eqnarray}
E_J(\Phi_{\rm ext})
= 2E_J \left|\cos\left(\pi\frac{\Phi_{\rm ext}}{\Phi_0}\right)\right|. 
\end{eqnarray}
For a discussion of the case with asymmetries in the SQUID parameters, see Ref.~\cite{Johansson2009}.%

%
%
So far, no assumptions have been made on the circuit parameters that determine the characteristic energy scales of the circuit, and both the waveguide fluxes and the SQUID flux are dynamical variables. However, from now on we assume that the plasma frequency of the SQUID far exceeds other characteristic frequencies in the circuit (e.g., the typical frequencies of the electromagnetic fields in the coplanar waveguide), so that oscillations in the phase across the SQUID have small amplitude, $\Phi_J/\Phi_0 \ll 1$, and the SQUID is operated in the phase regime, where $E_J(\Phi_{\rm ext}) \gg (2e)^2/2C_J$. The condition $\Phi_J/\Phi_0 \ll 1$ allows us to expand the cosine function in the SQUID Lagrangian, resulting in a quadratic Lagrangian
\begin{eqnarray}
\mathcal{L}
&=&
\frac{1}{2} \sum_{i=1}^{\infty} \left( \Delta x\,C_0(\dot\Phi_{i})^2-\frac{(\Phi_{i+1}-\Phi_{i})^2}{\Delta xL_0} \right) 
\nonumber\\
&+&
\frac{1}{2}C_{J}\,\dot\Phi_J^2 
-
\frac{1}{2}
\left(\frac{2\pi}{\Phi_0}\right)^2
E_{J}(\Phi_{\rm ext})\,\Phi_J^2.
\end{eqnarray}

Following the standard canonical quantization procedure, we can now transform the Lagrangian into a Hamiltonian which provides the quantum mechanical description of the circuit, using the Legendre transformation $H = \sum_i\frac{\partial \mathcal{L}}{\partial \dot\Phi_i}\dot\Phi_i - \mathcal{L}$. We obtain the following circuit Hamiltonian:
\begin{eqnarray}
H
&=&
\frac{1}{2} \sum_{i=1}^{\infty} \left(\frac{P_i^2}{\Delta xC_0}+\frac{(\Phi_{i+1}-\Phi_{i})^2}{\Delta xL_0} \right) 
\nonumber\\
&+&
\frac{1}{2}\frac{P_1^2}{C_J}
+
\frac{1}{2}
\left(\frac{2\pi}{\Phi_0}\right)^2
E_{J}(\Phi_{\rm ext})\,
\Phi_1^2,
\end{eqnarray}
and the commutation relations $[\Phi_i, P_j] = i\hbar\delta_{ij}$ and $[\Phi_i, \Phi_j] = [P_i, P_j] = 0$, where $P_j = \frac{\partial \mathcal{L}}{\partial \dot\Phi_j}$.
In the expression above we have also made the identification $\Phi_J \equiv \Phi_1$ (see Fig.~\ref{fig:cpw-squid-circuit-diagram}). The Heisenberg equation of motion for the flux operator $\Phi_1$ plays the role of a boundary condition for the field in the coplanar waveguide. 
By using the commutation relations given above, the equation of motion is found to be
\begin{eqnarray}
\dot P_1
&=& C_J\ddot\Phi_1 = -i[P_1, H] =
\nonumber\\
&=&  -\,E_J(\Phi_{\rm ext})\left(\frac{2\pi}{\Phi_0}\right)^2\Phi_1-\frac{1}{L_0}\frac{(\Phi_2-\Phi_1)}{\Delta x},\;\;
\end{eqnarray}
which in the continuum limit $\Delta x \rightarrow 0$ results in the boundary condition \cite{Wallquist2006}
\begin{eqnarray}
\label{eq:bnd-cont-cpw-squid}
C_J\,\ddot\Phi(0,t) &+& \left(\frac{2\pi}{\Phi_0}\right)^2\!\!E_J(t)\,\Phi(0,t)
\nonumber\\
&+&\frac{1}{L_0}\left.\frac{\partial\Phi(x,t)}{\partial x}\right|_{x=0} = 0,
\end{eqnarray}
where $\Phi_1(t) \equiv \Phi(x=0,t)$, and $E_J(t) = E_J[\Phi_{\rm ext}(t)]$.

This is the parametric boundary condition that can be tuned by the externally applied magnetic flux. Below we show how, under certain conditions, this boundary condition can be analogous to the boundary condition imposed by a perfect mirror at an effective length from the waveguide-SQUID boundary.

In a similar manner, we can derive the equation of motion for the dynamical fluxes in the coplanar waveguide (away from the boundary), i.e., for $\Phi_i$, $i>1$, which results in the well-known massless scalar Klein-Gordon equation. The general solution to this one-dimensional wave equation has independent components that propagate in opposite directions, and we identify these two components as the input and output components of the field in the coplanar waveguide.
 
\subsection{Quantization of the field in the waveguide}

Following e.g.~Refs.~\cite{Yurke1984,Devoret1995}, we now introduce creation and annihilation operators for the flux field in the coplanar waveguide,
and write the field in second quantized form:
\begin{eqnarray}
\label{eq:sec1-field}
\Phi(x,t)
 &=&
 \sqrt{\frac{\hbar Z_0}{4\pi}}
 \int_{0}^{\infty}\frac{d\omega}{\sqrt{\omega}}
 \left(
   a_{\rm in}(\omega) \; e^{-i(-k_\omega x+\omega t)} 
 \right. 
\nonumber\\
 &+& 
 \left.
   a_{\rm out}(\omega) \; e^{-i(k_\omega x+\omega t)} + \mathrm{H.c.}
 \right),
\end{eqnarray}
where $Z_0 = \sqrt{L_0/C_0}$ is the characteristic impedance.
We have separated the left- and right-propagating signals along the $x$-axis,
and denoted them as ``output'' and ``input'', respectively. The annihilation and creation operators satisfy the canonical commutation relation,
$[a_{\rm in(out)}(\omega'), a_{\rm in(out)}^\dagger(\omega'')] = \delta(\omega'-\omega'')$, and the wave vector is defined as $k_\omega = |\omega|/v$,
where $v = 1/\sqrt{C_0L_0}$ is the propagation velocity of photons in the waveguide.

Our goal is to characterize the output field, e.g., by calculating the expectation values and correlation functions of various combinations of output-field operators. 
To achieve this goal, we use the input-output formalism:
We substitute the expression for the field 
into the boundary condition imposed by the SQUID, 
and we solve for the output-field operators in terms of the input-field operators.
The input field is assumed to be in a known state, e.g., a thermal state or the vacuum state.

\subsection{Output field operators}

By substituting Eq.~(\ref{eq:sec1-field}) into the boundary condition, Eq.~(\ref{eq:bnd-cont-cpw-squid}), and Fourier transforming the result, we obtain a boundary condition in terms of the creation and annihilation operators (for $\omega' > 0$),
\begin{eqnarray}
\label{eq:sec-1-frequency-space-final-bnd}
0 
&=&
\left(\frac{2\pi}{\Phi_0}\right)^2 
\int_{-\infty}^{\infty}\!\!\!d\omega\,
g(\omega, \omega') \times
\nonumber\\
&& \left[
\Theta(\omega)(a^{\rm in}_\omega+a^{\rm out}_\omega) + \Theta(-\omega)(a^{\rm in}_{-\omega}+a^{\rm out}_{-\omega})^\dag
\right]
\nonumber\\
&-&
\omega'^2 C_J
(a^{\rm in}_{\omega'}+a^{\rm out}_{\omega'})
+
i\frac{k_{\omega'}}{L_0}
(a^{\rm in}_{\omega'}-a^{\rm out}_{\omega'}).
\end{eqnarray}
where
\begin{eqnarray}
  g(\omega, \omega') = \frac{1}{2\pi}\sqrt{\frac{|\omega'|}{|\omega|}} \int_{-\infty}^{\infty}\!\!\!dt\, E_J(t) e^{-i(\omega-\omega') t}.
\end{eqnarray}
This equation cannot be solved easily in general, but below we consider two cases where we can solve it analytically, i.e., when $E_J(t)$ is ($i$) constant, or ($ii$) has an harmonic time-dependence. In the general case, we can only solve the equation numerically, see Appendix \ref{app:numerical-input-output}. In Sec.~\ref{sec:measurement-setups} we compare the analytical results with such numerical calculations.

\subsubsection{Static applied magnetic flux}

If the applied magnetic flux is time-independent, $E_J(t) = E_J^0$, we obtain 
$g(\omega, \omega') = E_J^0 \sqrt{\left|\frac{\omega'}{\omega}\right|} \delta(\omega-\omega')$, 
and the solution takes the form
\begin{eqnarray}
\label{eq:a_out_static}
a_{\rm out}(\omega)
=
R(\omega)\;a_{\rm in}(\omega),
\end{eqnarray}
where
\begin{eqnarray}
\label{eq:R_full}
R(\omega)
&=&
-\,
\frac{
\left(\frac{2\pi}{\Phi_0}\right)^2  E_J^0
-
|\omega|^2 C_J
+
\frac{ik_{\omega}}{L_0}
}
{
\left(\frac{2\pi}{\Phi_0}\right)^2  E_J^0
-
|\omega|^2 C_J
-
\frac{ik_{\omega}}{L_0}
}.
\end{eqnarray}

\begin{figure}[t]
\includegraphics[width=8.5cm]{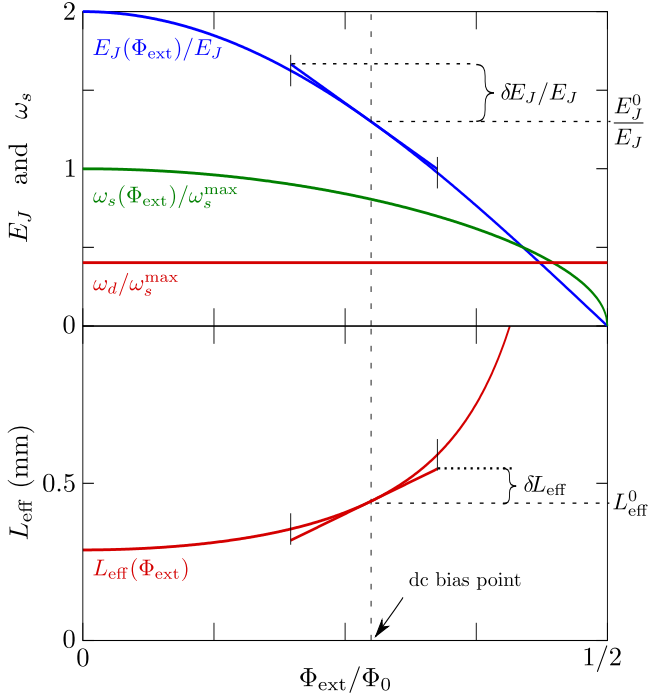}
\caption{(color online) The top panel shows the normalized effective Josephson energy, $E_J(\Phi_{\rm ext})$, and the normalized plasma frequency of the SQUID, $\omega_s(\Phi_{\rm ext})$, as a function of the external applied magnetic flux through the SQUID, $\Phi_{\rm ext}$. The driving frequency $\omega_d$, which should be much lower than the SQUID's plasma frequency, is also shown as a reference. The bottom panel shows the corresponding effective length, $L_{\rm eff}(\Phi_{\rm ext})$. The dashed vertical line marks the bias point used in the calculations, and the amplitude of the harmonic drive around this bias point is also indicated, by the linearized region. The parameters used in the calculations are, if nothing else is specified, $E_J^0 = 1.3 E_J$, $\delta\!E_J = E_J^0/4$, $E_J = I_c\Phi_0/(2\pi)$, where $I_c = 1.25\;\mu$A is the critical current of the Josephson junctions in the SQUID, $C_J=90$ fF, $v=1.2\times10^8$ m/s, $Z_0 \approx 55\;\Omega$, and $\omega_s = 37.3$ GHz, $\omega_d = 18.6$ GHz. These parameters result in an effective length $L_{\rm eff}^0 = 0.44$ mm, and an effective-length modulation $\delta\!L_{\rm eff} =  0.11$ mm.}
\label{fig:plot_Eeff_and_Leff}
\end{figure}

Assuming that the $|\omega|^2C_J$ term is small compared to the other terms, i.e., that the SQUID plasma frequency is sufficiently large, we can neglect it in the expression above, and we are left with the following simplified form:
\begin{eqnarray}
\label{eq:R_simplified}
R(\omega)\; = \; - \frac{1 + ik_\omega L_{\rm eff}^0}{1 - ik_\omega L_{\rm eff}^0}
\approx
- \exp\!\left\{2ik_\omega L_{\rm eff}^0\right\}.
\end{eqnarray}
Here, we have defined
\begin{eqnarray}
L_{\rm eff}^0 =  \left(\frac{\Phi_0}{2\pi}\right)^2\!\frac{1}{E_J^0L_0},
\end{eqnarray}
and assumed that $k_\omega L_{\rm eff}^0 \ll 1$ (this condition gives an upper bound on the frequencies for which this treatment is valid). 
Figure~\ref{fig:plot_Eeff_and_Leff} shows the dependences of $E_J$ and $L_{\rm eff}$ on the externally applied magnetic flux $\Phi_{\rm ext}$.

The reflection coefficient $R(\omega)$ on the simplified form given above [Eq.~(\ref{eq:R_simplified})],
exactly coincides with the reflection coefficient, $-\exp\{2ik_\omega L\}$, of a short-circuited coplanar waveguide of length $L$.
It is therefore natural to interpret the parameter $L_{\rm eff}^0$ as an effective length
that gives the distance from the SQUID to a perfectly reflecting mirror (which is equivalent to a short-circuit termination in the context of coplanar waveguides).
Alternatively, this can be phrased in terms of boundary conditions,
where the mixed-type boundary condition Eq.~(\ref{eq:bnd-cont-cpw-squid}) at $x=0$ is then equivalent to a Dirichlet boundary condition of an ideal mirror at $x=L_{\rm eff}^0$, for the frequencies satisfying $\omega \ll v/L_{\rm eff}^0$. See, e.g., Refs.~\cite{Alves2006,Alves2010} for discussions of different types of boundary conditions in the context of the dynamical Casimir effect.

\subsubsection{Weak harmonic drive}

For a weak harmonic applied magnetic flux
with frequency $\omega_d$, giving $E_J(t) = E_J^0 + \delta\!E_J \cos(\omega_d t)$, with $\delta\!E_J \ll E_J^0$, we obtain
\begin{eqnarray}
 g(\omega, \omega')
&=&
E_J^0 \sqrt{\frac{|\omega'|}{|\omega|}} \delta(\omega-\omega')
\nonumber\\
&+&
\delta\!E_J^0 \sqrt{\frac{|\omega'|}{|\omega|}} 
\frac{1}{2}
[
\delta(\omega-\omega'+\omega_d)
+
\delta(\omega-\omega'-\omega_d)
].
\nonumber\\
\end{eqnarray}
Inserting this into the boundary condition above, and assuming that $\omega' > 0$, we obtain (after renaming $\omega'\rightarrow\omega$)
\begin{eqnarray}
0 
&=&
(a^{\rm in}_{\omega}+a^{\rm out}_{\omega}) + ik_{\omega}L_{\rm eff}
(a^{\rm in}_{\omega}-a^{\rm out}_{\omega})
\nonumber\\
&+&
\frac{1}{2} \frac{\delta\!E_J^0}{E_J^0}
\left\{
\sqrt{\frac{\omega}{\omega-\omega_d}} 
\Theta(\omega-\omega_d)(a^{\rm in}_{\omega-\omega_d}+a^{\rm out}_{\omega-\omega_d})
\right.
\nonumber\\
&+&
\sqrt{\frac{\omega}{\omega_d-\omega}} 
\Theta(\omega_d-\omega)(a^{\rm in}_{\omega_d-\omega}+a^{\rm out}_{\omega_d-\omega})^\dag
\nonumber\\
&+&
\left.
\sqrt{\frac{\omega}{\omega+\omega_d}} 
(a^{\rm in}_{\omega+\omega_d}+a^{\rm out}_{\omega+\omega_d})
\right\},
\end{eqnarray}
where $\Theta(\omega)$ is the Heaviside step function,
and where we also here have assumed that the SQUID is in the ground state.
This equation cannot be solved exactly, because of the mixed-frequency terms (which creates an infinite series of sidebands around $\omega$), but we can take a perturbative approach assuming that $\frac{\delta\!E_J^0}{E_J^0} \ll 1$, which results in 
\begin{eqnarray}
\label{eq:a_out_as_a_in_raw}
&&a_{\rm out}(\omega)
=
R(\omega)\;a_{\rm in}(\omega) +
\nonumber\\
&&
S(\omega, \omega+\omega_d) e^{i(k_{\omega}+k_{\omega_d-\omega})L_{\rm eff}^0} a_{\rm in}(\omega + \omega_d) +
\nonumber\\
&&
S(\omega, \omega-\omega_d) e^{i(k_{\omega}+k_{\omega_d-\omega})L_{\rm eff}^0} a_{\rm in}(\omega-\omega_d) +
\nonumber\\
&&
S^*(\omega, \omega_d-\omega) e^{i(k_{\omega}-k_{\omega_d-\omega})L_{\rm eff}^0} a_{\rm in}^\dag(\omega_d-\omega),
\end{eqnarray}
where $R(\omega)$ is given by Eq.~(\ref{eq:R_simplified}), and
\begin{eqnarray}
\label{eq:S-omega-omega-d}
S(\omega', \omega'') = i\frac{\delta\!L_{\rm eff}}{v} \sqrt{\omega'\omega''}\, \Theta(\omega')\Theta(\omega''),
\end{eqnarray}
where $\delta\!L_{\rm eff} = L_{\rm eff}^0 \delta\!E_J / E_J^0$.
Here, 
\begin{eqnarray}
\label{eq:eps-single-mirror}
\epsilon = \mathrm{max}\left\{\left|S(\omega, \omega_d-\omega)\right|\right\} = \frac{\delta\!L_{\rm eff}}{v}\frac{\omega_d}{2}
\end{eqnarray}
is the small parameter in the perturbation calculation.
We note that Eq.~(\ref{eq:a_out_as_a_in_raw}) is simplified by translating it along the $x$-axis, from the point $x=0$ to the point $x=L_{\rm eff}^0$
(which is the position of the effective mirror),
\begin{eqnarray}
\label{eq:a_out_as_a_in}
a_{\rm out}(\omega)
&=&
-a_{\rm in}(\omega) 
+
S(\omega, \omega+\omega_d) a_{\rm in}(\omega + \omega_d) 
\nonumber\\
&+&
S(\omega, \omega-\omega_d) a_{\rm in}(\omega-\omega_d) 
\nonumber\\
&+&
S^*(\omega, \omega_d-\omega) a_{\rm in}^\dag(\omega_d-\omega).
\end{eqnarray}

It is the {\it last term} in Eq.~(\ref{eq:a_out_as_a_in}) that give rise to the {\it dynamical Casimir radiation in the output field} of the coplanar waveguide,
and it appears as a consequence of the {\it mixing} of the $a_{\rm in}$ and $a_{\rm in}^\dagger$ operators due to the time-dependent boundary condition.
Given this expression for $a_{\rm out}(\omega)$, we can in principle calculate any property of the output field.
In Sec.~\ref{sec:measurement-setups} we discuss a number of observables of the output field that
contain signatures of the presence of the dynamical Casimir part of the field described by the equations above.

\section{The dynamical Casimir effect in an open resonator circuit}
\label{sec:sc-dce-resonator}

\begin{figure}[t]
\includegraphics[width=8.0cm]{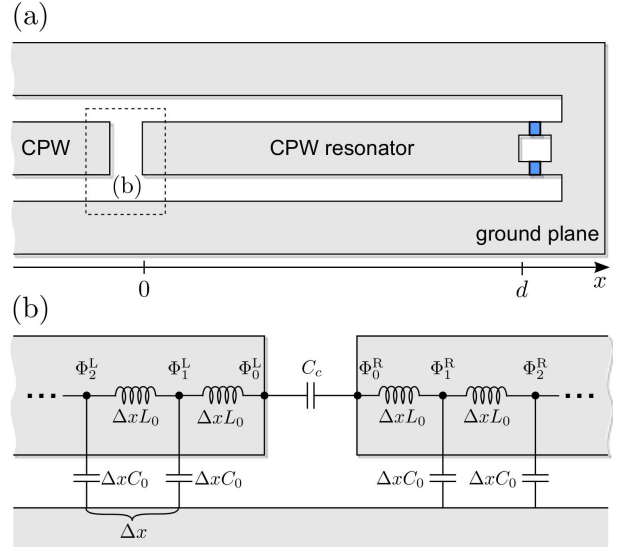}
\caption{(color online) (a) Schematic drawing of a coplanar waveguide (CPW) resonator, of length $d$, capacitively coupled to an open, semi-infinite coplanar waveguide to the left, and terminated to ground through a SQUID to the right. (b) A magnification of the capacitive gap between the resonator and the semi-infinite waveguide, shown as a dashed box in (a), together with its equivalent circuit model.}
\label{fig:drawing-cpw-res-squid}
\end{figure}

In the previous section we discussed a setup that corresponds to a single oscillating mirror in free space.
However, we note that experimentally it might be hard to completely avoid all resonances in the waveguide,
and in this section we therefore analyze the case where the waveguide is interrupted by a small gap at some
distance from the SQUID. Effectively this system forms an open coplanar waveguide resonator with time-dependent boundary condition,
where the size of the gap determines the coupling strength between the resonator and the waveguide. 
This system closely resembles a single-sided cavity in free space, with one oscillating mirror.
A schematic representation of the circuit is shown in Fig.~\ref{fig:drawing-cpw-res-squid}(a). This circuit has large parts in common with the circuit considered in Sec.~\ref{sec:sc-dce-single-mirror} (see Figs.~\ref{fig:cpw-squid-schematic} and \ref{fig:cpw-squid-circuit-diagram}). The new component is the capacitive gap that interrupts the coplanar waveguide, at $x = 0$, as shown in Fig.~\ref{fig:drawing-cpw-res-squid}(a) and (b). This gap is responsible for the formation of a resonator between the SQUID and the semi-infinite waveguide, at $x < 0$. The coupling strength between the resonator and waveguide determines the quality factor of the resonator. This quality factor, and the corresponding decay rate, are important parameters in the following analysis. Note that in the limit of vanishing quality factor this setup reduces to the setup studied in the previous section.

Figure \ref{fig:drawing-cpw-res-squid}(b) shows a lumped circuit model for the part of the circuit in the proximity of the capacitive gap.
The Lagrangian for this part of the circuit is 
\begin{eqnarray}
\mathcal{L} &=&
\frac{1}{2}
\Delta x C_0(\dot\Phi_{0}^{\rm L})^2
-
\frac{1}{2}
\frac{(\Phi_{1}^{\rm L}-\Phi_{0}^{\rm L})^2}{\Delta x\,L_0}
+
\frac{1}{2}
\Delta x C_0(\dot\Phi_{1}^{\rm L})^2
+ 
...
\nonumber\\
&+&
\frac{1}{2}
\Delta x C_0(\dot\Phi_{0}^{\rm R})^2
-
\frac{1}{2}
\frac{(\Phi_{1}^{\rm R}-\Phi_{0}^{\rm R})^2}{\Delta x\,L_0}
+
\frac{1}{2}
\Delta x C_0(\dot\Phi_{1}^{\rm R})^2
+
...
\nonumber\\ 
&+&
\frac{1}{2} C_{c}(\dot\Phi_{0}^{\rm L}-\dot\Phi_{0}^{\rm R})^2,
\end{eqnarray}
where $\Phi^{\rm L}_i$ and $\Phi^{\rm R}_i$ are the flux fields to the left and right of the capacitive gap, respectively.
In the continuum limit, where $\Delta x \rightarrow 0$, the equations of motion for $\Phi_0^{\rm L}$ and $\Phi_0^{\rm R}$
result in the following boundary condition for the field in the coplanar waveguide on both sides of the gap:
\begin{eqnarray}
\label{eq:tl_tlr_bnd_eq_1}
-\frac{1}{L_0}\left.\frac{\partial\Phi^{\rm L}(x,t)}{\partial x}\right|_{x=0^-}
\!&=&\!
C_c\left[
\left.\frac{\partial^2\Phi^{\rm L}}{\partial t^2}\right|_{x=0^-} -
\left.\frac{\partial^2\Phi^{\rm R}}{\partial t^2}\right|_{x=0^+}
\right]\nonumber\\
\label{eq:tl_tlr_bnd_eq_2}
\frac{1}{L_0}\left.\frac{\partial\Phi^{\rm R}(x,t)}{\partial x}\right|_{x=0^+}
\!&=&\!
C_c\left[
\left.\frac{\partial^2\Phi^{\rm R}}{\partial t^2}\right|_{x=0^+} - 
\left.\frac{\partial^2\Phi^{\rm L}}{\partial t^2}\right|_{x=0^-}
\right].
\nonumber\\
\end{eqnarray}
Using the field quantization from Sec.~\ref{sec:sc-dce-single-mirror}, and Fourier transforming the boundary condition above, results in a boundary condition in terms of creation and annihilation operators, in the frequency domain:
\begin{eqnarray}
&-&
\frac{ik_\omega}{L_0}
 \left[
   a_{\rm in}^{\rm L}(\omega)
-
   a_{\rm out}^{\rm L}(\omega)  
 \right]
=
\nonumber\\
&-&
\omega^2 C_c
 \left[
   a_{\rm in}^{\rm L}(\omega)
+
   a_{\rm out}^{\rm L}(\omega)
-
   a_{\rm in}^{\rm R}(\omega) 
-
   a_{\rm out}^{\rm R}(\omega) 
 \right]
\\
&&
\frac{ik_\omega}{L_0}
 \left[
   a_{\rm in}^{\rm R}(\omega)
-
   a_{\rm out}^{\rm R}(\omega)  
 \right]
=
\nonumber\\
&-&
\omega^2 C_c
 \left[
   a_{\rm in}^{\rm R}(\omega)
+
   a_{\rm out}^{\rm R}(\omega)
-
   a_{\rm in}^{\rm L}(\omega) 
-
   a_{\rm out}^{\rm L}(\omega) 
 \right]
\end{eqnarray}
This system of equations can be solved, and the operators for the resonator can be written in terms of the operators of the semi-infinite coplanar waveguide
\begin{eqnarray}
\label{eq:transformation_R_L}
\begin{pmatrix} 
 a_{\rm  in}^{\rm R}(\omega) \\
 a_{\rm out}^{\rm R}(\omega) \\
\end{pmatrix}
&=&
\begin{pmatrix} 
  1 - i\frac{\omega_c}{2\omega} & i\frac{\omega_c}{2\omega} \\
  -i\frac{\omega_c}{2\omega}    & 1+i\frac{\omega_c}{2\omega} \\
\end{pmatrix} 
\begin{pmatrix} 
 a_{\rm  in}^{\rm L}(\omega) \\
 a_{\rm out}^{\rm L}(\omega) \\
\end{pmatrix},
\end{eqnarray}
where $\omega_c = (C_cZ_0)^{-1}$ is a parameter that characterizes the coupling strength between the resonator and the open coplanar waveguide.
This transformation
can be used together with Eq.~(\ref{eq:a_out_as_a_in}), which relates the input and output operators of the field in direct contact with the effective mirror
(i.e., $a^{\rm R}_{\rm in}(\omega, x=d_{\rm eff})$ and $a^{\rm R}_{\rm out}(\omega, x=d_{\rm eff})$, in the present setup),
to obtain a relation between $a^{\rm L}_{\rm in}(\omega)$ and $a^{\rm L}_{\rm out}(\omega)$ that do not contain the resonator operators.
To achieve this we note that the resonator operators at $x=0$ are related to those at $x=d_{\rm eff}$ by a simple phase factor,
according to the transformation
\begin{eqnarray}
\label{eq:a_translation}
\!
\begin{pmatrix} 
 a_{\rm  in}^{\rm R}(\omega, 0) \\
 a_{\rm out}^{\rm R}(\omega, 0) \\
\end{pmatrix}
\!=\!
\begin{pmatrix} 
  e^{ik_\omega d_{\rm eff}} & 0                          \\
  0                         & e^{-ik_\omega d_{\rm eff}} \\
\end{pmatrix}\!\!
\begin{pmatrix} 
 a_{\rm  in}^{\rm R}(\omega, d_{\rm eff}) \\
 a_{\rm out}^{\rm R}(\omega, d_{\rm eff}) \\
\end{pmatrix}\!\!,\;\;\;\;\;
\end{eqnarray}
and that $a^{\rm R}_{\rm out}(\omega,0)$ and $a^{\rm R}_{\rm in}(\omega,0)$ are related to $a^{\rm L}_{\rm out}(\omega)$ and $a^{\rm L}_{\rm in}(\omega)$ according to the transformation in Eq.~(\ref{eq:transformation_R_L}).

\subsection{Output field operators}

\subsubsection{Static magnetic flux}

The resonator boundary condition on the side that is terminated by the SQUID is described by Eq.~(\ref{eq:a_out_as_a_in}). 
In the case of a static applied magnetic field the inelastic scattering by the effective mirror is absent, 
i.e., $S(\omega', \omega'') = 0$, and only the elastic reflections remain,
\begin{eqnarray}
 a^{R}_{\rm out}(\omega, d_{\rm eff}) = -\,a^{R}_{\rm in}(\omega, d_{\rm eff}).
\end{eqnarray} 
In the present setup, the SQUID is located at $x=d$, and the effective mirror is located at $x=d_{\rm eff}$, 
where $d_{\rm eff}=d+L_{\rm eff}^0$.
Thus, to write a relation between the input and output operator that applies on the left side of the resonator,
we translate the boundary condition of the effective mirror by $d_{\rm eff}$,
according to Eq.~(\ref{eq:a_translation}),
\begin{eqnarray}
\!a^{R}_{\rm out}(\omega, 0) \exp\{-ik_\omega d_{\rm eff}\} = - a^{R}_{\rm in}(\omega, 0) \exp\{ik_\omega d_{\rm eff}\}.\;\;\;
\end{eqnarray}
Since this relation applies at the point where the resonator is capacitively coupled to the open coplanar waveguide,
we can transform it using Eq.~(\ref{eq:transformation_R_L}),
\begin{eqnarray}
&&\left[
-i\frac{\omega_c}{2\omega}a^{L}_{\rm in}(\omega)
+
\left(1+i\frac{\omega_c}{2\omega}\right)a^{L}_{\rm out}(\omega)
\right]\exp\{-2ik_\omega d_{\rm eff}\}
\nonumber\\
&=& 
-
\left[
\left(1 - i\frac{\omega_c}{2\omega}\right)a^{L}_{\rm in}(\omega)
+
i\frac{\omega_c}{2\omega}a^{L,0}_{\rm out}(\omega)
\right].
\end{eqnarray}
This equation can be rewritten as
\begin{eqnarray}
a^{L}_{\rm out}(\omega) = R_{\rm res}(\omega)a^{L}_{\rm in}(\omega), 
\end{eqnarray}
where 
\begin{eqnarray}
\label{eq:R_res_full}
R_{\rm res}(\omega)
&=&
\frac
{1+\left(1+2i\omega/\omega_c\right)\exp\{2ik_\omega d_{\rm eff}\}}
{\left(1-2i\omega/\omega_c\right) + \exp\{2ik_\omega d_{\rm eff}\}}.
\end{eqnarray}

\begin{figure}[t]
\includegraphics[width=8.5cm]{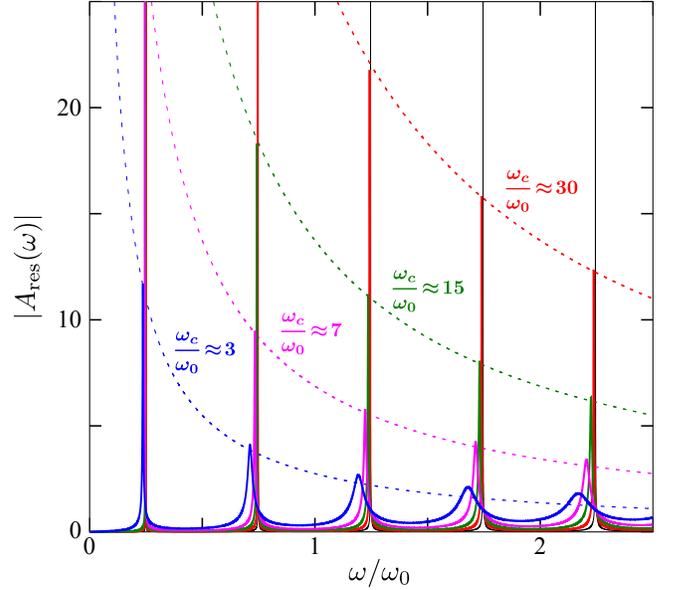}
\caption{(color online) The absolute value of $A_{\rm res}(\omega)$ as a function of the renomalized frequency $\omega/\omega_0$,
where $\omega_0 = 2\pi v/d_{\rm eff}$ is the full-wavelength resonance frequency when the resonator is decoupled from the open coplanar waveguide.
The sequence of curves correspond to different coupling strengths to the CPW,
characterized by resonator $\omega_c$ values indicated by the labels in the figure.}
\label{fig:plot-resonator-modes}
\end{figure}

Similarly, we can apply Eq.~(\ref{eq:transformation_R_L}) to solve for the resonator operators in terms of the input operators for the coplanar waveguide,
\begin{eqnarray}
a^{R}_{\rm out}(\omega, d_{\rm eff}) = A_{\rm res}(\omega)a^{L}_{\rm in}(\omega), 
\end{eqnarray}
where
\begin{eqnarray}
A_{\rm res}(\omega)
&=&
\frac
{(2i\omega/\omega_c) \exp\{ik_\omega d_{\rm eff}\}}
{\left(1-2i\omega/\omega_c\right) + \exp\{2ik_\omega d_{\rm eff}\}},
\end{eqnarray}
The function $A_{\rm res}(\omega)$ describes the resonator's response to an input signal from the coplanar waveguide,
and it contains information about the mode structure of the resonator.
From $A_{\rm res}(\omega)$ we can extract the resonance frequencies and the quality factors for each mode, see Fig.~\ref{fig:plot-resonator-modes}.

The resonance frequencies, $\omega^{\rm res}_n$, are approximately given by the transcendental equation
\begin{eqnarray}
\tan\left(2\pi\omega^{\rm res}_n/\omega_0\right) &=& \omega_c/\omega^{\rm res}_n,
\end{eqnarray}
where $\omega_0=2\pi v/d_{\rm eff}$ and $n$ is the mode number. The corresponding resonance widths and quality factors are
\begin{eqnarray}
\label{eq:resonator-G-n}
\Gamma_n 
&=& 2 \frac{\omega_0}{2\pi} \left( \frac{\omega^{\rm res}_n}{\omega_c} \right)^2,
\\
\label{eq:resonator-Q-n}
Q_n 
&\equiv& \frac{\omega^{\rm res}_n}{\Gamma_n} 
= 2\pi\frac{\omega_c^2}{2\omega_0\omega^{\rm res}_n},
\end{eqnarray}
respectively. We note that the quality factor for higher-order modes are rapidly decreasing as a function of the mode number $n$ (see also Fig.~\ref{fig:plot-resonator-modes}).

Using the expressions for $\omega_n^{\rm res}$ and $\Gamma_n$ given above,
the resonator response can be expanded around the resonance frequencies and written in the form
\begin{eqnarray}
\label{eq:resonator-breit-wigner-form}
A_{\rm res}(\omega) \approx -\sqrt{\frac{\omega_0}{2\pi}}\frac{\sqrt{\Gamma_n/2}}{\Gamma_n/2 - i(\omega - \omega^{\rm res}_n)},
\end{eqnarray}
and, similarly, the expression for the reflection coefficient of the resonator from the open coplanar waveguide is
\begin{eqnarray}
 R_{\rm res}(\omega) \approx - \frac{\Gamma_n/2 + i(\omega - \omega^{\rm res}_n)}{\Gamma_n/2 - i(\omega - \omega^{\rm res}_n)}.
\end{eqnarray}

\subsubsection{Weak harmonic drive}

For a time-dependent applied magnetic flux in the form of a weak harmonic drive, we again take a perturbative approach and solve the equations for $a^{\rm L}_{\rm out}(\omega)$ in terms of $a^{\rm L}_{\rm in}(\omega)$ by treating $S(\omega', \omega'')$ as a small parameter. Following the approach of the previous Section, 
we eliminate the resonator variables by using Eq.~(\ref{eq:a_out_as_a_in}) and Eq.~(\ref{eq:transformation_R_L}), and we obtain
\begin{eqnarray}
\label{eq:a_out_as_a_in_resonator}
a_{\rm out}^{\rm L}(\omega)
&=&
R_{\rm res}(\omega)\;a_{\rm in}^{\rm L}(\omega)
\nonumber\\
&+&
S_{\rm res,1}(\omega, \omega+\omega_d) a_{\rm in}^{\rm L}(\omega + \omega_d)
\nonumber\\
&+&
S_{\rm res,1}(\omega, \omega-\omega_d) a_{\rm in}^{\rm L}(\omega-\omega_d)
\nonumber\\
&+&
S_{\rm res,2}^*(\omega, \omega_d-\omega) (a_{\rm in}^{\rm L})^\dag(\omega_d-\omega),\;\;
\end{eqnarray}
where $R_{\rm res}(\omega)$ is given by Eq.~(\ref{eq:R_res_full}), and
\begin{eqnarray}
\label{eq:S-res-omega-omega-d}
S_{\rm res,1}(\omega', \omega'')
&=&
S(\omega',\omega'')
A_{\rm res}(\omega')
A_{\rm res}(\omega''),
\\
\label{eq:S-res-omega-omega-d-II}
S_{\rm res,2}(\omega', \omega'')
&=&
S(\omega',\omega'')
A_{\rm res}^*(\omega')
A_{\rm res}(\omega'').
\end{eqnarray}
In this case the small parameter in the perturbation calculation is
\begin{eqnarray}
\label{eq:eps-resonator}
\epsilon_{\rm res} = 
\mathrm{max}\left\{\left|S_{\rm res,2}(\omega,\omega_d-\omega)\right|\right\} = 
\frac{\delta\!L_{\rm eff}}{d_{\rm eff}}\frac{\omega_d}{2}\frac{1}{\Gamma_n}.\;\;
\end{eqnarray}
We note that Eq.~(\ref{eq:a_out_as_a_in_resonator}) has the same general form as Eq.~(\ref{eq:a_out_as_a_in}), and that the only differences are the definitions of the reflection and inelastic scattering functions: $R_{\rm res}(\omega)$ and $S_{\rm res,\alpha}(\omega', \omega'')$. This similarity allows us to analyze both cases using the same formalism in the following sections, where we calculate expectation values and correlation functions of physically relevant combinations of the output field operators.

\section{Measurement setups}
\label{sec:measurement-setups}

Equation (\ref{eq:a_out_as_a_in}) in Sec.~\ref{sec:sc-dce-single-mirror}, and Eq.~(\ref{eq:a_out_as_a_in_resonator}) in Sec.~\ref{sec:sc-dce-resonator}, constitute complete theoretical descriptions of the corresponding output fields, and we can apply these expressions in calculating the expectation values of any output-field observable or correlation function. In this Section we discuss possible experimental setups for measuring various physical properties of the output field, and we discuss which quantum mechanical observables and correlation functions these setups measure, in terms of the output-field operators. Below, we use Eq.~(\ref{eq:a_out_as_a_in}) and Eq.~(\ref{eq:a_out_as_a_in_resonator}), and explicitly evaluate these physical observables for the two setups discussed in the previous Sections.
\begin{figure}[t]
\includegraphics[width=8.0cm]{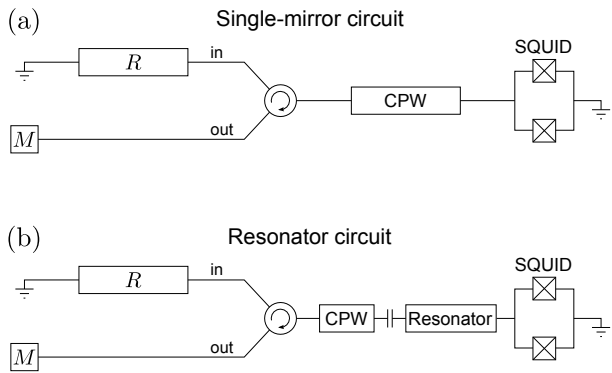}
\caption{Schematic diagrams of the measurement setups for (a) the single-mirror and (b) the resonator setups in superconducting microwave circuits, as discussed in Sec.~\ref{sec:sc-dce-single-mirror} and Sec.~\ref{sec:sc-dce-resonator}, respectively. The circulator (indicated by a circle with a curved arrow) separates the input and output fields such that only the signal from the SQUID reaches the measurement device $M$,
and so that the input-field state is given by the thermal Johnson-Nyquist noise from the resistive load $R$, at temperature $T$.}
\label{fig:schematic-mw-circuit}
\end{figure}

The basic setups that we are considering here are shown schematically in Fig.~\ref{fig:schematic-mw-circuit}, which also illustrates the concept of separating the incoming and the outgoing fields by means of a circulator. The input field is terminated to ground through an impedance-matched resistive load. This resistor produces a Johnson-Nyquist (thermal) noise that acts as the input on the SQUID. The circulator isolates the detector from the thermal signal from the resistor, except for the part of the noise that is reflected on the SQUID. The measurement device is denoted by $M$ in these circuits.

We are interested in experimentally relevant observables that contain signatures of the dynamical Casimir part of the output field, 
i.e., the part that is described by the fourth term in Eq.~(\ref{eq:a_out_as_a_in}) and Eq.~(\ref{eq:a_out_as_a_in_resonator}). 
The most distinct signature of the dynamical Casimir effect is perhaps the correlations between individual pairs of photons,
and such correlations could in principle be measured in a coincidence-count experiment.
However, the physical quantities that can be measured in a microwave circuit are slightly different from those measured in the quantum optics regime.
For instance, there are currently no single-photon detectors available in the microwave regime,
and as a consequence it is not possible to directly measure the correlations between individual pairs of photons.
Instead, there are linear amplifiers \cite{Bozyigit2010} that can amplify weak signals, with very low-intensity photon-flux densities, to larger signals that can be further processed with classical electronics. In addition to amplifying the signal, these amplifiers also add noise \cite{Caves1982,Clerk2010} to the signal. However, the increased noise can often be compensated for, e.g., by averaging the signal over long a period of time, or by measuring cross-correlations in which the noise mostly cancel out.

\begin{figure}[t]
\includegraphics[width=6.6cm]{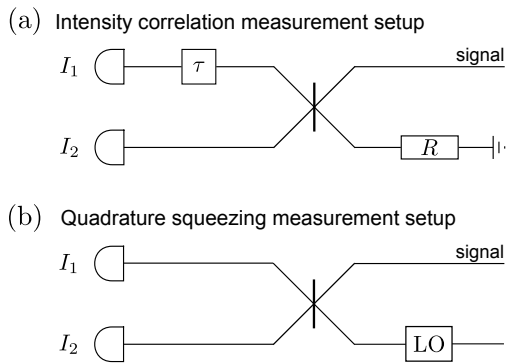}
\caption{Schematic measurement setups for (a) intensity correlations and for (b) quadrature squeezing.
In (a), the box with $\tau$ inside represents a time-delay, and in (b) the box with the label LO represents a local oscillator.
The microwave beam splitter can be implemented, e.g., by a hybrid ring. The detectors are assumed to measure the intensity of the voltage field.
In a practical experimental setup, the signals would also have to pass through several stages of amplification, which are not shown here.}
\label{fig:schematic-mw-detectors}
\end{figure}

In microwave electronics, the natural fields for describing the coplanar waveguides are the current and voltage fields, 
and these physical quantities can be readily measured with standard equipment.
Here we therefore focus on the voltage $V(x,t)$ in the coplanar waveguide as our main physical observable.
The voltage is related to the previously defined field creation and annihilation operators according to
\begin{eqnarray}
\label{eq:voltage-quantum}
 &&V_{\rm out}(x,t)
 =
\partial_t \Phi_{\rm out}(x,t) =
\nonumber\\
 &&\sqrt{\frac{\hbar Z_0}{4\pi}}
 \int_{0}^{\infty}d\omega \sqrt{\omega}
 \left(
   - i a^{\rm out}_\omega \; e^{-i(k_\omega x+\omega t)} + \mathrm{H.c.}
 \right).
\end{eqnarray}
The output field states described by Eq.~(\ref{eq:a_out_as_a_in}) and Eq.~(\ref{eq:a_out_as_a_in_resonator}) have voltage expectation values that are zero,
$\left<V_{\rm out}(x,t)\right>=0$, 
but the squared voltages, i.e., as measured by a voltage square-law detector,
and various forms of voltage correlations,
can have non-zero expectation values.
For example, 
$\left<V_{\rm out}(x,t)^2\right>$,
$\left<V_{\rm out}(t_1)V_{\rm out}(t_2)\right>$, and
$\left<V_{\rm out}(\omega_1)V_{\rm out}(\omega_2)\right>$,
are all in general non-zero,
and do contain {\it signatures of the presence of the dynamical Casimir radiation}.

\subsection{Photon-flux density}

To measure the photon-flux density requires an intensity detector, such as a photon counter that clicks each time a photon is absorbed by the detector. The measured signal is proportional to the rate at which the detector absorbs photons from the field, which in turn is proportional to the field intensity. This detector model is common in quantum optics,
and it is also applicable to intensity detectors (such as voltage square-law detectors) in the microwave regime, although not with single-photon resolution.

The signal recorded by a quantum mechanical photon intensity (power) detector (see, e.g., Ref.~\cite{Glauber1963}) in the coplanar waveguide corresponds to the observable
\begin{eqnarray}
 I(t) \propto {\rm Tr}\left[\rho \hat{V}^{(-)}(x,t)\hat{V}^{(+)}(x,t)\right],
\end{eqnarray}
where $V^{(\pm)}(x,t)$ are the positive and negative frequency component of the voltage field, respectively. In terms of the creation and annihilation operators for the field in the coplanar waveguide [see Eq.~(\ref{eq:voltage-quantum})], where we, for brevity, have taken $x=0$, 
\begin{eqnarray}
 I(t) \propto 
\int_{0}^{\infty}\!\!\!d\omega'
\int_{0}^{\infty}\!\!\!d\omega'' \sqrt{\omega'\omega''}\,
{\rm Tr}
\left[
\rho
a_{\omega'}^\dag
a_{\omega''}
\right]
e^{i(\omega'-\omega'')  t},
\nonumber\\
\end{eqnarray}
and the corresponding noise-power spectrum is
\begin{eqnarray}
\mathcal{S}_V(\omega) 
&=&
\int_0^\infty\!\!\! d\omega'\,
{\rm Tr}\left[\rho \hat{V}^{(-)}(\omega)\hat{V}^{(+)}(\omega')\right]\nonumber\\
&=&\frac{\hbar Z_0}{4\pi}  
\int_0^\infty\!\!\! d\omega'\,
\sqrt{\omega\omega'}\; n(\omega, \omega'),
\end{eqnarray}
where $n(\omega, \omega') = \mathrm{Tr}\left[\rho a^\dagger(\omega)a(\omega)\right]$.
Here, $\mathcal{S}_V(\omega)$ is 
related to the voltage auto-correlation function 
via a Fourier transform, according to the Wiener-Khinchin theorem.
The photon-flux density in the output field, 
\begin{eqnarray}
 n_{\rm out}(\omega) = \int_0^{\infty}\!\!\!d\omega'\;n_{\rm out}(\omega,\omega'),
\end{eqnarray}
can be straightforwardly evaluated using Eq.~(\ref{eq:a_out_as_a_in}). The resulting expression is
\begin{eqnarray}
\label{eq:nout_w}
n_{\rm out}(\omega)
&=&
|R(\omega)|^2
\bar{n}_{\rm in}(\omega)
+
|S(\omega, \omega + \omega_d)|^2
\bar{n}_{\rm in}(|\omega+\omega_d|)
\nonumber\\
&+&
|S(\omega, |\omega-\omega_d|)|^2
\bar{n}_{\rm in}(|\omega-\omega_d|)
\nonumber\\
&+&
|S(\omega, \omega_d-\omega)|^2
\Theta(\omega_d-\omega),
\end{eqnarray}
where 
$\bar{n}_{\rm in}(\omega) =  {\rm Tr}\left[\rho a_{\rm in}^\dag(\omega) a_{\rm in}(\omega)\right]$
is the thermal photon occupation of the input-field mode with frequency $\omega$, 
given by $\bar{n}^{\rm in}_\omega = [\exp(\hbar\omega/k_bT)-1]^{-1}$, 
where $T$ is the temperature and $k_B$ is the Boltzmann constant.

\begin{figure}[t]
\includegraphics[width=8.5cm]{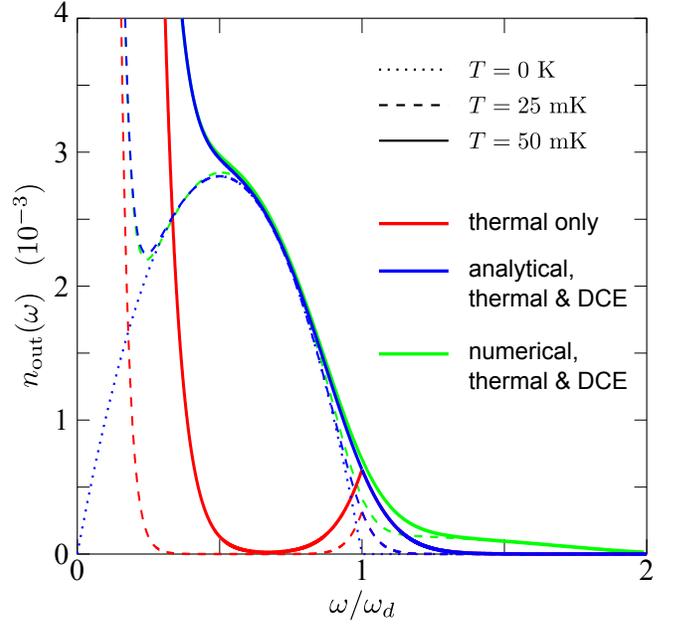}
\caption{(color online) The output-field photon-flux density, $n_{\rm out}(\omega)$, as a function of the relative mode frequency $\omega/\omega_d$, for the single-mirror setup. The solid and the dashed curves are for the temperatures $T=50$ mK and $T=25$ mK, respectively, and the dotted curve is for zero temperature.
The red (bottom) curves show the part of the signal with thermal origin, and the blue (dark) and the green (light) curves also include the radiation due to the dynamical Casimir effect.
The blue curves are the analytical results,
and the green curves are calculated numerically using the method described in Appendix \ref{app:numerical-input-output}.
The parameters are the same as in Fig.~\ref{fig:plot_Eeff_and_Leff}. The presence of the dynamical Casimir radiation is clearly distinguishable for temperatures up to $\sim 70$ mK. The good agreement between the analytical and numerical results verifies the validity of the perturbative calculation for the
parameters used here.
}
\label{fig:photon-flux-tl-squid}
\end{figure}

The first three terms in the expression above are of thermal origin, and the fourth term is due to the dynamical Casimir effect. In order for the dynamical Casimir effect not to be negligible compared to the thermally excited photons, we require that $k_bT \ll \hbar\omega_d$, where the driving frequency $\omega_d$ here serves as a characteristic frequency for the system, since all dynamical Casimir radiation occurs below this frequency (to leading order). In this case it is safe to neglect the term containing the small factor $\bar{n}_{\rm in}(|\omega+\omega_d|)$ in the expression above. Substituting the expression for $S(\omega,\omega_d)$, from Eq.~(\ref{eq:S-omega-omega-d}), into Eq.~(\ref{eq:nout_w}), results in the following explicit expression for the output field photon flux, for the single-mirror case:
\begin{eqnarray}
\label{eq:nout_w_squid}
n_{\rm out}(\omega)
=
\bar{n}_{\rm in}(\omega)
&+&
\left(\frac{\delta\!L_{\rm eff}}{v}\right)^2
\!\!\omega\;|\omega-\omega_d|\;
\bar{n}_{\rm in}(|\omega-\omega_d|)
\nonumber\\
&+&
\left(\frac{\delta\!L_{\rm eff}}{v}\right)^2
\!\!\omega(\omega_d-\omega)\;
\Theta(\omega_d-\omega).\;\;\;\;
\end{eqnarray}

The output-field photon flux density, Eq.~(\ref{eq:nout_w_squid}), is plotted in Fig.~\ref{fig:photon-flux-tl-squid}. The blue dotted parabolic contribution to $n_{\rm out}(\omega)$ is due to the fourth term in Eq.~(\ref{eq:nout_w}), i.e., the dynamical Casimir radiation [compare Eq.~(\ref{eq:review_n_w})].

\begin{figure}[t]
\includegraphics[width=8.5cm]{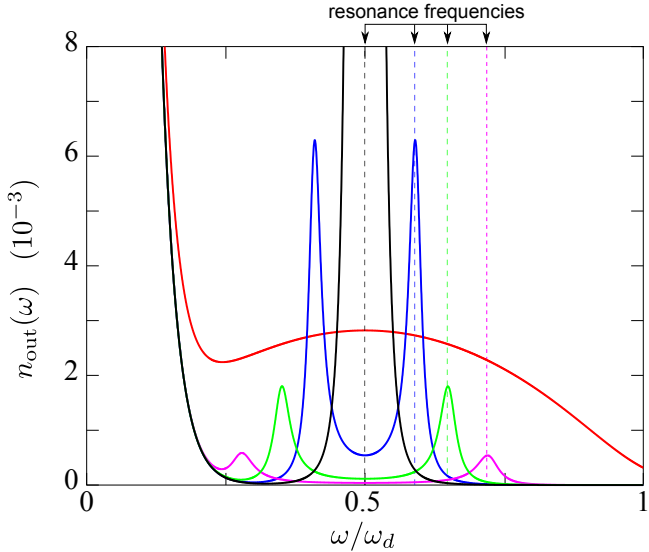}
\caption{(color online) The output photon-flux density, $n_{\rm out}(\omega)$, for the resonator setup (solid lines), as a function of the normalized frequency $\omega/\omega_d$,
for four difference resonance frequencies (marked by dashed vertical lines). Note that a double-peak structure appears when the driving frequency is detuned from twice the resonance frequency. For reference, the result for the single-mirror setup is also shown (red solid curve without resonances). Here, the temperature was chosen to be $T = 25$ mK, and the quality factor of the first resonance mode is $Q_0 \approx 20$ ($\omega_c \approx 3 \omega_d$), see Eq.~(\ref{eq:resonator-Q-n}).
The other parameters are the same as in Fig.~\ref{fig:plot_Eeff_and_Leff}.}
\label{fig:plot-photon-flux-tl-res-squid-finite-temperature}
\end{figure}
\begin{figure}[t]
\includegraphics[width=8.5cm]{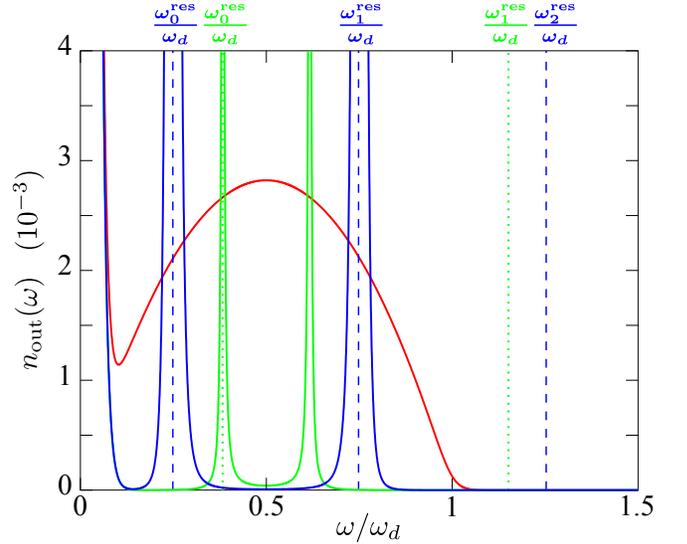}
\caption{(color online) The output photon-flux density, $n_{\rm out}(\omega)$, as a function of normalized frequency,
for the resonator setups where two modes (blue) and a single mode (green) are active in the dynamical Casimir radiation.
The dashed and the dotted vertical lines mark the resonance frequencies of the few lowest modes for the two cases, respectively. The solid red curve shows the photon-flux density in the absence of the resonator. 
The blue solid curve is the photon-flux density for the two-mode resonance, i.e., for the case when the two lowest resonance frequencies add up to the driving frequency. The green curve shows the photon-flux density for the case when only a single mode in the resonator is active (see Fig.~\ref{fig:plot-photon-flux-tl-res-squid-finite-temperature} for more examples of this case).
Here, the temperature was chosen to be $T = 10$ mK, and the resonator's quality factor is in both cases $Q_0 \approx 50$.
The other parameters are the same as in Fig.~\ref{fig:plot_Eeff_and_Leff}.}
\label{fig:plot-photon-flux-res-squid-two-vs-one-mode-resonance}
\end{figure}

Similarly, by using Eq.~(\ref{eq:a_out_as_a_in_resonator}), we calculate the output-field photon-flux density for the setup with a resonator, and the resulting expression also takes the form of Eq.~(\ref{eq:nout_w}), but where $R(\omega)$ and $S(\omega', \omega'')$ are given by Eq.~(\ref{eq:R_res_full}) and Eq.~(\ref{eq:S-res-omega-omega-d}), respectively. The resulting photon-flux density for the resonator setup is plotted in Fig.~\ref{fig:plot-photon-flux-tl-res-squid-finite-temperature}. Here, photon generation occurs predominately in the resonant modes of the resonator. For significant dynamical Casimir radiation to be generated it is necessary that the frequencies of both generated photons ($\omega'$ and $\omega''$, where $\omega'+\omega''=\omega_d$) 
are near the resonant modes of the resonator. In the special case when the first resonance coincide with half of the driving frequency, $\omega^{\rm res}_n = \omega_d/2$, there is a resonantly-enhanced emission from the resonator, see Fig.~\ref{fig:plot-photon-flux-tl-res-squid-finite-temperature}.
The resonant enhancement is due to parametric amplification of the electric field in the resonator (i.e., amplification of both thermal photons and photons generated from vacuum fluctuations due to the dynamical Casimir effect). 

Another possible resonance condition is
\begin{eqnarray}
\omega^{\rm res}_0+\omega^{\rm res}_1 \sim \omega_d.
\end{eqnarray}
In this case, strong emission can occur even 
when the frequencies of the two generated photons are significantly different,
since the two photons can be resonant with different modes of the resonator, i.e., $\omega' \sim \omega^{\rm res}_0$ and $\omega'' \sim \omega^{\rm res}_1$.
See the blue curves in Fig.~\ref{fig:plot-photon-flux-res-squid-two-vs-one-mode-resonance}.

We conclude that in the case of a {\it waveguide without any resonances the observation of the parabolic shape of the photon-flux density $n_{\rm out}(\omega)$ would be a clear signature of the dynamical Casimir effect}. 
The parabolic shape of the photon-flux density should also be distinguishable in the presence of a realistic thermal noise.
Resonances 
in the waveguide concentrate the photon-flux density to the vicinity of the resonance frequencies, which can give a larger signal with a smaller bandwidth.
One should note that in order to stay in the perturbative regime, the driving amplitude should be reduced by the quality factor of the resonance, 
compared to the case without any resonances.
The bimodal structure of the spectrum, and its characteristic behavior as a function of the driving frequency and detuning, should be a clear indication of the dynamical Casimir effect. 

\subsection{Two-photon correlations}

\begin{figure}[t]
\includegraphics[width=8.5cm]{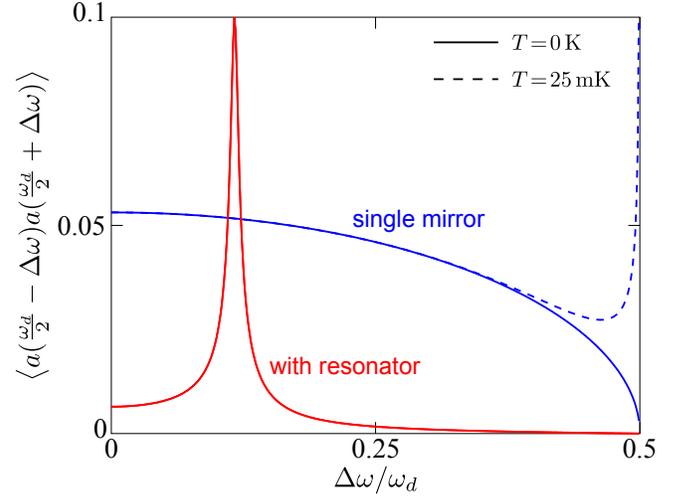}
\caption{(color online) Correlations between photons at the frequencies $\omega_d/2-\Delta\omega$ and $\omega_d/2+\Delta\omega$, as a function of detuning $\Delta\omega$ from half the driving frequency, for the single-mirror setup (blue) and the setup with a resonator (red) that is slightly detuned from half the driving frequency, $\omega_d/2$. The dashed blue curve shows the correlations for single-mirror setup in the presence of thermal noise, at $T = 25$ mK.}
\label{fig:plot-aout-aout}
\end{figure}

The output fields described by Eq.~(\ref{eq:a_out_as_a_in}) and Eq.~(\ref{eq:a_out_as_a_in_resonator}) exhibit correlations between photons at different frequencies. This is straightforwardly demonstrated by calculating the expectation value of the photon-annihilation operators at two frequencies symmetric around half the driving frequency, i.e., 
\begin{eqnarray}
\label{eq:a_out_a_out}
\left< a_{\rm out}\left(\frac{\omega_d}{2}-\Delta\omega\right)a_{\rm out}\left(\frac{\omega_d}{2}+\Delta\omega\right)\right>
=
\nonumber\\
R\left(\frac{\omega_d}{2}-\Delta\omega\right)
S^*\left(\frac{\omega_d}{2}+\Delta\omega, \frac{\omega_d}{2}-\Delta\omega\right)\times
\nonumber\\
\left[
1
+
\bar{n}_{\rm in}\left(\frac{\omega_d}{2}-\Delta\omega\right)
\right],
\end{eqnarray}
which can be interpreted as the correlation (entanglement) between two photons
that are simultaneously created at the frequencies $\omega_d/2-\Delta\omega$ and $\omega_d/2+\Delta\omega$,
where $\Delta\omega < \omega_d/2$.
This two-photon correlation is shown in Fig.~\ref{fig:plot-aout-aout},
for the field generated by the SQUID without the resonator (shown in blue), and with the resonator (shown in red).
Note that for thermal and vacuum states this expectation value vanishes for all frequencies $\omega$.
This correlation is not directly measurable, since the operator combination is not Hermitian,
but it serves the purpose of being the most basic illustration of the presence of nonclassical two-photon correlations
in the field produced by the dynamical Casimir effect.
Below we consider two physically observable correlation functions that are experimentally measurable.

\subsubsection{Second-order coherence function}

\begin{figure}[t]
\includegraphics[width=8.5cm]{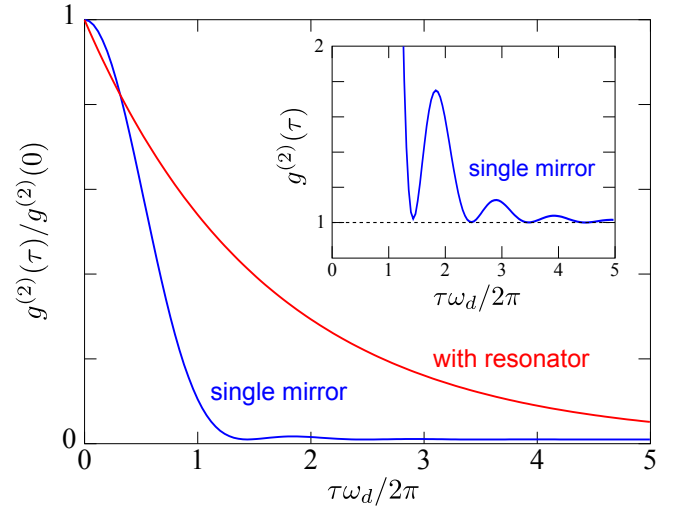}
\caption{The normalized second-order coherence function, $g_2(\tau)$, as a function of the delay time $\tau$,
for the field produced by the single-mirror setup from Sec.~\ref{sec:sc-dce-single-mirror} (in blue), 
and for the resonator setup (in red), at zero temperature.
The inset shows the second-order coherence function for the single-mirror setup without the normalization,
illustrating that $g_2(\tau) > 1$ and showing the oscillating behavior of $g_2(\tau)$ for large $\tau$.}
\label{fig:plot-intensity-correlations}
\end{figure}

The fact that  photons are predicted to be generated in pairs in the dynamical Casimir effect
implies that the time-domain photon statistics exhibits photon bunching.
For instance, the measurement setup outlined in Fig.~\ref{fig:schematic-mw-detectors}(a),
which measures the second-order correlation function
\begin{eqnarray}
G^{(2)}(\tau)
&=&
\mathrm{Tr}\left[\rho V^{(-)}(0)V^{(-)}(\tau)V^{(+)}(\tau)V^{(+)}(0)\right]\!,\;\;\;\;
\end{eqnarray}
could be used to detect this bunching effect.
For the output fields on the form of Eqs.~(\ref{eq:a_out_as_a_in}) and (\ref{eq:a_out_as_a_in_resonator}),
this correlation function takes the form
\begin{eqnarray}
&&G^{(2)}(\tau)
=
|G^{(1)}(\tau)|^2
+
\left|\int_0^{\omega_d}\!\!\!d\omega\,\omega\,|S(\omega , \omega_d - \omega)|^2e^{i\omega\tau}\right|^2
\nonumber\\
&&
+
\left|
\int_0^{\omega_d}
\!\!\!d\omega
\sqrt{\omega(\omega_d-\omega)}
R(\omega_d-\omega)S^*(\omega,\omega_d-\omega)e^{i\omega\tau}
\right|^2,\nonumber\\
&&
G^{(1)}(\tau)
=\int_0^{\omega_d}\!\!\!d\omega\,\omega |S(\omega , \omega_d - \omega)|^2,
\end{eqnarray}
where $R(\omega)$ and $S(\omega',\omega'')$ are defined
by Eqs.~(\ref{eq:R_full},\ref{eq:S-omega-omega-d}) and (\ref{eq:R_res_full},\ref{eq:S-res-omega-omega-d-II}) for the two setups, respectively.
The normalized second-order correlation function, i.e.,
the second-order coherence function
\begin{eqnarray}
 g^{(2)}(\tau) = \frac{G^{(2)}(\tau)}{G^{(1)}(0)G^{(1)}(\tau)},
\end{eqnarray}
is shown in Fig.~\ref{fig:plot-intensity-correlations} for the single-mirror setup (blue) and the resonator setup (red).
These coherence functions show clear photon bunching, since $g^{(2)}(\tau) > 1$ for a large range in $\tau$.
In particular, for zero time-delay, $\tau = 0$, the coherence functions can be written as
\begin{eqnarray}
 g^{(2)}(0) = 2 + \frac{1}{\epsilon^2},
\end{eqnarray}
where $\epsilon$ is given by Eq.~(\ref{eq:eps-single-mirror}) for the single-mirror setup,
and by Eq.~(\ref{eq:eps-resonator}) for the resonator setup, 
as discussed in Sec.~\ref{sec:sc-dce-single-mirror} and Sec.~\ref{sec:sc-dce-resonator}, respectively.
In both cases, $\epsilon$ is small and $g^{(2)}(0)\gg1$, which corresponds to large photon bunching.
The high value of $g^{(2)}(0)$ can be understood from the fact that the photons are created in pairs, so the probability to detect two photons simultaneously is basically the same as the probability to detect one photon. For low photon intensities, this gives a very large second-order coherence. The decay of $g^{(2)}(\tau)$ is given by the bandwidth of the photons, which in the case without resonance is given by the driving frequency $\omega_d$. When a resonance is present, its bandwidth $\Gamma$ determines the decay. Squeezed states show this type of photon bunching \cite{Loudon1987}, and we now proceed to calculate the squeezing spectrum of the radiation. 

\subsubsection{Squeezing spectrum}

Another nonclassical manifestation of the pairwise photon correlation in the fields described by Eqs.~(\ref{eq:a_out_as_a_in}) and (\ref{eq:a_out_as_a_in_resonator}) is quadrature squeezing \cite{Caves1985,Zagoskin2008} and the corresponding squeezing spectrum \cite{Walls1994},
defined as the quadrature squeezing at a certain frequency.
The quadratures in the frequency domain are defined by the relation 
\begin{eqnarray}
 X_\theta(\omega) = \frac{1}{2}\left[a(\omega)e^{-i\theta}+a^\dag(\omega)e^{i\theta}\right],
\end{eqnarray}
so that $X_1 = X_{\theta = 0}$, and $X_2 = X_{\theta = \pi/2}$.
Experimentally, the quadratures in a continuous multimode field can be measured through homodyne detection,
where the signal field
is mixed with a local oscillator (LO) on a balanced beam splitter, resulting in
$a_{\rm out}(t) = (a_{\rm LO}(t) + a_{\rm sig}(t)) / \sqrt 2$.
See Fig.~\ref{fig:schematic-mw-detectors}(b) for a schematic representation of this setup.
The local oscillator field is assumed to be in a large-amplitude coherent state with frequency $\Omega$ and phase $\theta$, i.e., 
$a_{\rm LO} = |\alpha|\exp\{-i(\theta+\Omega t)\}$.
Probing the resulting output field with an intensity detector then provides information about the quadrature in the signal field, since
\begin{eqnarray}
I(t) &=& \left<a_{\rm out}^\dag(t) a_{\rm out}(t)\right> 
\approx
|\alpha|^2 + |\alpha|
\left<X_{\rm out}^\theta(t)\right>
\end{eqnarray}
where
\begin{eqnarray}
 X_{\rm out}^\theta(t) &=& \frac{1}{2}\left[a_{\rm sig}(t)e^{i(\theta+\Omega t)} + a_{\rm sig}^\dag(t) e^{-i(\theta+\Omega t)}\right].
\end{eqnarray}
The noise-power spectrum of the voltage intensity of the output field therefore gives the squeezing spectrum of the signal field, in the frame rotating with frequency $\Omega$:
\begin{eqnarray}
\mathcal{S}_X^\theta(\Delta\omega) = 1\!+\!4\!\int_{-\infty}^{\infty}\!\!\!dt \,e^{-i\Delta\omega t} \left<: \Delta\!X^\theta_{\rm out}(t) \Delta\!X^\theta_{\rm out}(0):\right>,\nonumber\\
\end{eqnarray}
where $\left<:\,\,\,:\right>$ is the normally-ordered expectation value, and
where we have normalized the squeezing spectrum so that $\mathcal{S}_X^\theta = 1$ for unsqueezed vacuum,
and $\mathcal{S}_X^\theta = 0$ corresponds to maximum squeezing.
Here, $\Delta\omega$ is the frequency being measured after the mixing with the local oscillator,
and it is related to the frequency $\omega$ in the signal field as $\omega = \Omega + \Delta\omega$.
Hereafter, we choose $\Omega = \frac{\omega_d}{2}$.
\begin{figure}[t]
\includegraphics[width=8.5cm]{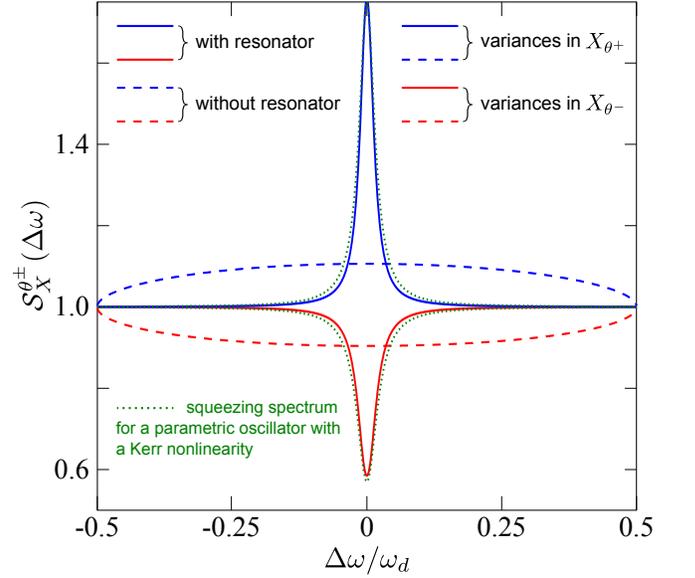}
\caption{(color online) The spectra of quadrature squeezing in the output field for a SQUID-terminated coplanar waveguide with (solid lines) and without (dashed lines) a resonator, as a function of the renormalized frequency detuning from $\omega_d/2$. The blue (dark) and the red (light) lines correspond to the variances in the $X_{\theta^-}$ and $X_{\theta^+}$ quadrature, respectively. For reference, the dotted thin lines show the squeezing spectrum for the field produced by a parametric oscillator with a Kerr nonlinearity.}
\label{fig:plot-squeezing-spectrum}
\end{figure}

Evaluating the squeezing spectrum for the quadrature defined by the relative phase $\theta$, i.e., $X_{\rm out}^\theta(t)$, results in
\begin{eqnarray}
\label{eq:squeezing_spectrum_general}
&&\mathcal{S}_{X}^\theta(\Delta\omega) = 
1 + 
2\left|S\left(\frac{\omega_d}{2}+\Delta\omega,\frac{\omega_d}{2}-\Delta\omega\right)\right|^2
\nonumber\\
&+& 
e^{-2i\theta} R  \left(\frac{\omega_d}{2}+\Delta\omega\right) S  \left(\frac{\omega_d}{2}+\Delta\omega, \frac{\omega_d}{2}-\Delta\omega\right)
\nonumber\\
&+& 
e^{2i\theta}  R^*\left(\frac{\omega_d}{2}-\Delta\omega\right) S^*\left(\frac{\omega_d}{2}-\Delta\omega, \frac{\omega_d}{2}+\Delta\omega\right)
\end{eqnarray}
where, as before, $R(\omega)$ and $S(\omega',\omega'')$ are defined by Eqs.~(\ref{eq:R_full},\ref{eq:S-omega-omega-d}) and Eqs.~(\ref{eq:R_res_full},\ref{eq:S-res-omega-omega-d}) for the two setups, respectively.

For the single-mirror setup discussed in Sec.~\ref{sec:sc-dce-single-mirror}, we obtain the following squeezing spectrum
\begin{eqnarray}
\label{eq:squeezing_spectrum_single_mirror}
\mathcal{S}_{X}^{\theta}(\Delta\omega)
&\approx& 
1
-
2\epsilon \sin(2\theta)
\sqrt{1 - 4\left(\frac{\Delta\omega}{\omega_d}\right)^2},
\end{eqnarray}
where we have neglected the second term in Eq.~(\ref{eq:squeezing_spectrum_general}), which is one order higher in the small parameter $S(\omega',\omega'')$.
Here,
we can identify the relative phases $\theta^- = \pi/4$ and $\theta^+ = -\pi/4$
as the maximally-squeezed quadrature ($\theta^-$) and the corresponding orthogonal quadrature ($\theta^+$).

By using the expressions for reflection and inelastic scattering of the resonator setup, Eqs.~(\ref{eq:R_res_full},\ref{eq:S-res-omega-omega-d}),
in the expression for the squeezing spectrum, Eq.~(\ref{eq:squeezing_spectrum_general}), we obtain
\begin{eqnarray}
\label{eq:squeezing_spectrum_resonator}
&&\mathcal{S}_{X}^{\theta^\pm}(\Delta\omega) \approx 
1 
\pm
\frac{2\epsilon_{\rm res}}{1 + \left(2\frac{\Delta\omega}{\Gamma}\right)^2},
\end{eqnarray}
where the frequency of the first resonator mode is assumed to coincide with half the driving frequency, $\omega^{\rm res}_0 = \omega_d/2$,
and where we again have defined $\theta^\pm = \mp\pi/4$ to correspond to the maximally-squeezed quadrature and the corresponding orthogonal quadrature. 
The squeezing spectra  for the single-mirror setup, Eq.~(\ref{eq:squeezing_spectrum_single_mirror}),
and for the resonator setup, Eq.~(\ref{eq:squeezing_spectrum_resonator}),
are plotted in Fig.~\ref{fig:plot-squeezing-spectrum}. 
The squeezing is limited by $\epsilon$ and $\epsilon_{\rm res}$, respectively,
and it is therefore not possible to achieve perfect squeezing, 
but as shown in Fig.~\ref{fig:plot-squeezing-spectrum}, significant squeezing is still possible.
In the single-mirror case the squeezing covers a large bandwidth,
and the total squeezing (see, e.g., Ref.~\cite{Walls1994}) of the $X_{\theta^-}$ quadrature,
given by the integral of $\mathcal{S}_{X}(\omega, \theta^-)$, is
\begin{eqnarray}
 \mathcal{S}_{X}^{\rm Total}(\theta^-) \;\approx\; \omega_d\left(1 - \frac{\pi}{2}\epsilon\right).
\end{eqnarray}

\section{Comparison with a parametric oscillator}
\label{sec:parametric-oscillator}

\begin{table*}[tbp]
\begin{center}
\begin{tabular}{c|c|c|c}
\hline\hline
\parbox{3cm}{\vspace{0.2cm}\vspace{0.2cm}} & 
\parbox{4.5cm}{\vspace{0.2cm} Single-mirror DCE  \vspace{0.2cm}} & 
\parbox{4.5cm}{\vspace{0.2cm} Low-Q resonator DCE  \vspace{0.2cm}} & 
\parbox{4.5cm}{\vspace{0.2cm} High-Q resonator DCE / PO \vspace{0.2cm}} 
\\ \hline\hline
\parbox{3cm}{\vspace{0.2cm} Comments \vspace{0.2cm}} & 
\parbox{4.5cm}{\vspace{0.2cm} Photons created due to time-dependent boundary condition. \vspace{0.2cm}} & 
\parbox{4.5cm}{\vspace{0.2cm} The resonator slightly alters the mode density, compared to the single-mirror case. \vspace{0.2cm}} & 
\parbox{4.5cm}{\vspace{0.2cm} DCE in a high-Q resonator is equivalent to a PO below threshold. \vspace{0.2cm}}
\\ \hline
\parbox{3cm}{\vspace{0.2cm} Classical analogue?  \vspace{0.2cm}} & 
\parbox{4.5cm}{\vspace{0.2cm} No, requires vacuum fluctuations. \vspace{0.2cm}} & 
\parbox{4.5cm}{\vspace{0.2cm} No, requires vacuum fluctuations. \vspace{0.2cm}} & 
\parbox{4.5cm}{\vspace{0.2cm} Yes, vacuum and thermal fluctuations give similar results. \vspace{0.2cm}}
\\ \hline
\parbox{3cm}{\vspace{0.2cm} Resonance condition  \vspace{0.2cm}} & 
\parbox{4.5cm}{\vspace{0.2cm} no resonator \vspace{0.2cm}} & 
\parbox{4.5cm}{\vspace{0.2cm} $\omega_{\rm res} = \omega_d/2$ \vspace{0.2cm}} & 
\parbox{4.5cm}{\vspace{0.2cm} $\omega_{\rm res} = \omega_d/2$ \vspace{0.2cm}}
\\ \hline
\parbox{3cm}{\vspace{0.2cm} Threshold condition  \vspace{0.2cm}} & 
\parbox{4.5cm}{\vspace{0.2cm} -- \vspace{0.2cm}} & 
\parbox{4.5cm}{\vspace{0.2cm} $\epsilon_{\rm res} \sim Q^{-1}$ \vspace{0.2cm}} & 
\parbox{4.5cm}{\vspace{0.2cm} $\epsilon_{\rm res} \sim Q^{-1} \ll 1$ \\ \vspace{0.1cm}
Above threshold: nonlinearity dominates behavior. \vspace{0.2cm}}
\\ \hline
\parbox{3cm}{\vspace{0.2cm}  Number of DCE photons per second  \vspace{0.2cm}} & 
\parbox{4.5cm}{\vspace{0.2cm} $\sim n(\omega_d/2)\, \omega_d$     \vspace{0.2cm}} & 
\parbox{4.5cm}{\vspace{0.2cm} $\sim n(\omega_{\rm res})\, \Gamma$ \vspace{0.2cm}} & 
\parbox{4.5cm}{\vspace{0.2cm} $\sim n(\omega_{\rm res})\, \Gamma$ \vspace{0.2cm}}
\\ \hline
\parbox{3cm}{\vspace{0.2cm} Spectrum \\ at $T = 0$ K \vspace{0.2cm}} & 
\parbox{4.5cm}{\vspace{0.2cm} 
\includegraphics[width=2.25cm]{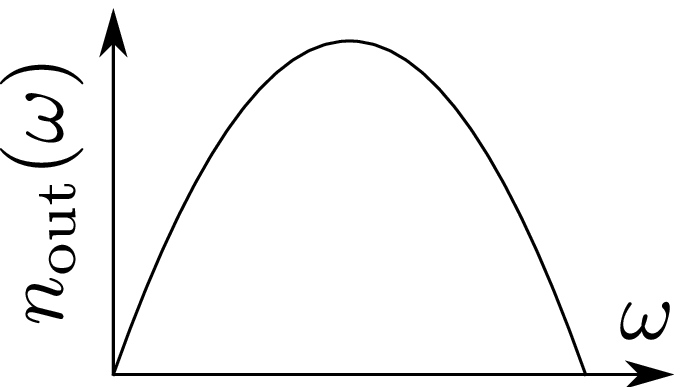} \\
Broadband spectrum with \\
peak at $\omega_d/2$ \vspace{0.2cm}} & 
\parbox{4.5cm}{\vspace{0.2cm} 
\includegraphics[width=4.5cm]{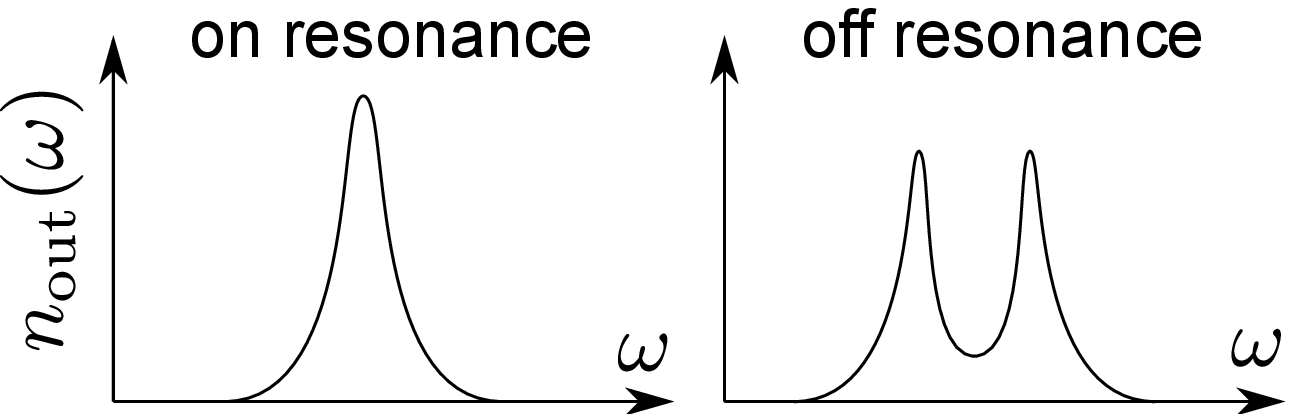} \\
Broad peaks at resonance frequency $\omega_{\rm res}$ and the complementary frequency $\omega_d-\omega_{\rm res}$. \vspace{0.2cm}} & 
\parbox{4.5cm}{\vspace{0.2cm}
\includegraphics[width=2.25cm]{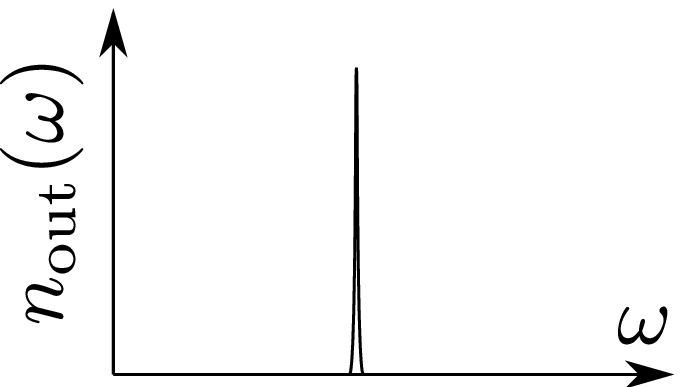} \\
Sharply peaked around the resonance frequency $\omega_{\rm res} = \omega_d/2$. \vspace{0.2cm}}
\\ \hline\hline
\end{tabular}
\end{center}
\caption{Comparison between the dynamical Casimir effect (DCE), in the single-mirror setup and the resonator setup, with a parametric oscillator (PO) with a
Kerr nonlinearity. In Sec.~\ref{sec:parametric-oscillator} we showed that in the high-$Q$ limit, the dynamical Casimir effect in the resonator setup is equivalent to a parametric oscillator. In the very-high-$Q$ limit, the dynamics involves many modes of the resonator. We do not consider the latter case here.}
\end{table*}

As the $Q_n$ values of the resonator considered in Sec.~\ref{sec:sc-dce-resonator} increases,
its resonant modes are increasingly decoupled from the coplanar waveguide,
and the modes become increasingly equidistant.
In the limit $Q_n\rightarrow\infty$,
the system formally reduces to the ideal case of a closed one-dimensional cavity
(see, e.g., Refs.~\cite{Moore1970,Dodonov1990,Dodonov1993,Law1994}).
However, this limit is not realistic for the type of circuits investigated here,
because it corresponds to a regime where also the high-frequency modes are significantly excited,
and this would violate
our assumption that the SQUID is adiabatic (i.e., that the SQUID plasma frequency is the largest frequency in the problem).
Our theoretical analysis is also unsuitable for studying that extreme limit,
since it implies that $\epsilon_{\rm res}$ no longer is small.

However, for moderate $Q_0$-values, where $\epsilon_{\rm res}$ is small and our analysis applies,
it is still possible to make a comparison to a single-mode parametric oscillator (PO) below its threshold.
The Hamiltonian for a pumped parametric oscillator \cite{Walls1994} with a Kerr-nonlinearity can be written as
\begin{eqnarray}
\!\!H_{\rm PO}
=
\frac{\hbar\omega_2}{2} a^\dag a
+
\frac{1}{2}i\hbar
\left[
e^{-i\omega_d t} \epsilon\left(a^\dag\right)^2
-
e^{i\omega_d t} \epsilon^* a^2
\right]\!\!,\;
\end{eqnarray}
and where the oscillator is assumed to couple to an environment that induces relaxation with a rate $\gamma$.
The output field for the parametric oscillator is described by
\begin{eqnarray}
\label{eq:a-out-po}
a_{\rm out}^{\rm PO}(\omega) = 
F(\omega) a_{\rm in}^{\rm PO}(\omega)
+
G(\omega) a_{\rm in}^{\rm PO}(- \omega)^\dag,
\end{eqnarray}
where
\begin{eqnarray}
F(\omega) &=& \frac{\left(\gamma/2\right)^2+\omega^2+|\epsilon|^2}{\left(\gamma/2-i\omega\right)^2 - |\epsilon|^2}, \\ 
G(\omega) &=& \frac{\gamma\epsilon}{\left(\gamma/2-i\omega\right)^2 - |\epsilon|^2},
\end{eqnarray}
see, e.g., Ref.~\cite{Walls1994}.
Comparing Eq.~(\ref{eq:a-out-po}) to the corresponding results for the dynamical Casimir effect:
\begin{eqnarray}
a_{\rm out}^{\rm DCE}(\omega) &=& 
R_{\rm res}\left(\frac{\omega_d}{2}+\omega\right) a_{\rm in}^{\rm DCE}(\omega)\nonumber\\
&+&
S_{\rm res}^*\left(\frac{\omega_d}{2}-\omega,\frac{\omega_d}{2}+\omega\right) a_{\rm in}^{\rm DCE}(- \omega)^\dag,
\end{eqnarray}
where $R_{\rm res}(\omega)$ and $S_{\rm res}(\omega',\omega'')$ are given by Eqs.~(\ref{eq:R_res_full},\ref{eq:S-res-omega-omega-d-II}),
allows us to identify relations between the parametric oscillator parameters (to first order in $\epsilon$) and the dynamical Casimir parameters.
We obtain
\begin{eqnarray}
 \gamma &=& \Gamma_0, \\
 \epsilon &=& -i \frac{\delta\!L_{\rm eff}\,\omega_d}{4\,d_{\rm eff}},
\end{eqnarray}
and thereby establish a one-to-one mapping between these systems,
valid for sufficiently large $Q$ and below the parametric oscillator threshold: $\epsilon < \gamma/2$, i.e.,
for 
\begin{eqnarray}
 \frac{\delta\!L_{\rm eff}}{d_{\rm eff}}\frac{\omega_d}{2\Gamma} < 1.
\end{eqnarray}
Using these expressions we can write a Hamiltonian that describes the dynamical Casimir effect in the resonator setup,
\begin{eqnarray}
H_{\rm DCE}
=
\frac{\hbar\omega_d}{2}a^\dag a
-
\frac{\delta\!L_{\rm eff}}{4d_{\rm eff}}
\frac{\hbar\omega_d}{2}
\left[
e^{ i\omega_d t} a^2
+ 
e^{-i\omega_d t}
\left(a^\dag\right)^2
\right],
\nonumber\\
\end{eqnarray}
and where the capacitive coupling to the open coplanar waveguide induces relaxation with a rate $\Gamma_0$ in the resonator.
This Hamiltonian picture
offers an alternative description of the photon creation process in the dynamical Casimir effect in a resonator.
This correspondence between the dynamical Casimir effect and a parametric oscillator was also discussed in e.g.~Ref.~\cite{Dezael2009}.

\section{Summary and conclusions}
\label{sec:summary}

We have analyzed the dynamical Casimir radiation in superconducting electrical circuits based on coplanar waveguides with tunable boundary conditions, which are realized by terminating the waveguides with SQUIDs. We studied the case of a semi-infinite coplanar wave\-guide, and the case of a coplanar waveguide resonator coupled to a semi-infinite waveguide, and we calculated the photon flux, the second-order coherence functions and the noise-power spectrum of field quadratures (i.e., the squeezing spectrum) for the radiation generated due to the dynamical Casimir effect.
These quantities have distinct signatures which can be used to identify the dynamical Casimir radiation in experiments.

For the single-mirror setup, we conclude that the photon-flux density $n_{\rm out}(\omega)$ has a distinct inverted parabolic shape that would be a clear signature of the dynamical Casimir effect. This feature in the photon-flux density should also be distinguishable in the presence of a realistic thermal noise background.

For the resonator setup, the presence of resonances in the coplanar waveguide alters the mode density and concentrates the photon-flux density,
of the dynamical Casimir radiation, around the resonances, which can result in a larger signal within a smaller bandwidth.
If the driving signal is detuned from the resonance frequency,
the resulting photon-flux density spectrum features a bimodal structure,
owing to the fact that photons are created in pairs with frequency that add up to the driving frequency.
The characteristic behavior of these features in the photon-flux density spectrum should also be a clear indication of the dynamical Casimir radiation. 
A resonance with a small quality factor could therefore make the experimental detection of the dynamical Casimir effect easier.
In the limit of large quality factor, however, the output field generated due to the dynamical Casimir effect becomes increasingly similar to that of a classical system, which makes it harder to experimentally identify the presence of the dynamical Casimir radiation \cite{Wilson2010}.

For both the single-mirror setup and the resonator setup with low quality factor,
the second-order coherence functions and the quadrature squeezing spectrum show 
signatures of the pairwise photon production and the closely related quadrature squeezing in the output field. 
The pairwise photon production of the dynamical Casimir effect has much in common with a parametrically driven oscillator,
and in the presence of a resonance this correspondence can be quantified, and the two systems can be mapped to each other
even though the systems have distinct physical origins.
This correspondence offers an alternative formulation of the dynamical Casimir effect in terms of a Hamiltonian for a resonator that is pumped via a nonlinear medium.

\begin{acknowledgments}
We would like to thank S.~Ashhab and N.~Lambert for useful discussions.
GJ acknowledges partial support by the European Commission through the
IST-015708 EuroSQIP integrated project and by the Swedish Research
Council.
FN acknowledges partial support from the 
Laboratory of Physical Sciences, 
National Security Agency, Army Research Office, 
National Science Foundation grant No. 0726909,
JSPS-RFBR contract No. 09-02-92114, 
Grant-in-Aid for Scientific Research (S), 
MEXT Kakenhi on Quantum Cybernetics, and 
Funding Program for Innovative R\&D on S\&T (FIRST).
\end{acknowledgments}

\appendix
\section{Numerical calculations of output field expectation values in the input-output formalism}
\label{app:numerical-input-output}

In this section we describe the methods applied in the numerical calculations of the expectation values and correlation functions of the output field.
Instead of taking a perturbative approach and solving for the output field operators in terms of the input field operators analytically,
we can solve the linear integral equation Eq.~(\ref{eq:sec-1-frequency-space-final-bnd}) numerically 
by truncating the frequency range to $[-\Omega, \Omega]$ 
and discretizing it in $(2N+1)$ steps $[-\omega_{N}, ..., \omega_0 = 0, ..., \omega_N]$, 
so that $\omega_N = \Omega$. 
Here it is also convenient to define $a(-\omega) = a^\dag(\omega)$, so that
the boundary condition in the frequency domain reads
\begin{eqnarray}
0 
&=&
\left(\frac{2\pi}{\Phi_0}\right)^2 
\int_{-\Omega}^{\Omega}\!\!d\omega
\left[
a^{\rm in}_\omega+a^{\rm out}_\omega
\right]
g(\omega, \omega')
\nonumber\\
&-&
|\omega'|^2 C_J
(a^{\rm in}_{\omega'}+a^{\rm out}_{\omega'})
+
\frac{i|\omega'|}{vL_0}
(a^{\rm in}_{\omega'}-a^{\rm out}_{\omega'}),
\end{eqnarray}
and in the discretized frequency space takes the form
\begin{eqnarray}
&&\sum_{m = -N}^N
\left[
-
\left(\frac{2\pi}{\Phi_0}\right)^2 
\!\!\Delta\omega\,
g({\omega_m}, \omega_n)
\right.
\nonumber\\
&&\left.
+
\frac{i|\omega_n|}{vL_0}
\delta_{\omega_n,\omega_m}
+
|\omega_n|^2 C_J
\delta_{\omega_n,\omega_m}
\right]
a^{\rm out}_{\omega_m}
\nonumber\\ 
&=&
\sum_{m = -N}^N
\left[
\left(\frac{2\pi}{\Phi_0}\right)^2 
\!\!\Delta\omega\,
g({\omega_m}, \omega_n)
\right.
\nonumber\\
&&\left.
+
\frac{i|\omega_n|}{vL_0}
\delta_{\omega_m,\omega_n}
-
|\omega_n|^2 C_J
\delta_{\omega_m,\omega_n}
\right]
a^{\rm in}_{\omega_m}
\end{eqnarray}
where we have substituted $\omega'\rightarrow\omega_n$ and $\omega\rightarrow\omega_m$.
This equation can be written in the matrix form
\begin{eqnarray}
 M_{\rm out} a_{\rm out} = M_{\rm in} a_{\rm in}
\,&\Rightarrow&\,
 a_{\rm out} = M_{\rm out}^{-1} M_{\rm in} a_{\rm in},
\end{eqnarray}
where
\begin{eqnarray}
 M^{\rm out}_{mn} = &-&
\left(\frac{2\pi}{\Phi_0}\right)^2
\!\Delta\omega\,
g({\omega_m}, \omega_n)
+
\frac{i|\omega_n|}{vL_0}
\delta_{\omega_n,\omega_m}
\nonumber\\
&+&
|\omega_n|^2 C_J
\delta_{\omega_n,\omega_m}
\\
 M^{\rm in}_{mn} = & & 
\left(\frac{2\pi}{\Phi_0}\right)^2 
\!\Delta\omega\,
g({\omega_m}, \omega_n)
-
\frac{i|\omega_n|}{vL_0}
\delta_{\omega_n,\omega_m}
\nonumber\\
&+&
|\omega_n|^2 C_J
\delta_{\omega_n,\omega_m}
\end{eqnarray}
and
\begin{eqnarray}
a^{\rm out}_{m} &=& \left(a^{\rm out}(\omega_{-N}), ...,a^{\rm out}(\omega_{0}), ..., a^{\rm out}(\omega_{N}) \right)^T
\\
a^{\rm in}_{m}  &=& \left(a^{\rm in}(\omega_{-N}), ...,a^{\rm in}(\omega_{0}), ..., a^{\rm in}(\omega_{N}) \right)^T
\end{eqnarray}
and, finally, where
\begin{eqnarray}
g(\omega_{m},\omega_{n})
= \frac{1}{2\pi}
\sqrt{\frac{|\omega_n|}{|\omega_m|}} \int_{-\infty}^{\infty}\!\!\!dt\,E_J(t) e^{-i(\omega_m-\omega_n) t},\nonumber\\
\end{eqnarray}
which can be obtained by a Fourier transform of the drive signal $E_J(t)$.

For an harmonic drive signal we can use the fact that the time-dependence in the boundary condition
only mixes frequencies that are integer multiples of the driving frequency,
and by selecting only these sideband frequencies in the frequency-domain expansion, 
i.e., $\omega_n = \omega + n\omega_d$ and $n = -N, ..., N$,
we obtain results that are more accurate than the perturbation results, if $N>1$.


\vspace*{-0.2in}
\begin{thebibliography}{99}
\vspace*{-0.2in}

\bibitem{Moore1970}
G.T. Moore, 
J. Math. Phys. {\bf 11} 2679 (1970).

\bibitem{Fulling1976}
S.A. Fulling and P.C.W. Davies, 
Proc. R. Soc. London, Ser. A {\bf 348}, 393 (1976).

\bibitem{Barton1993}
G.~Barton and C.~Eberlein,
Ann. Phys. {\bf 227}, 222 (1993).

\bibitem{Dodonov2001}
V.V.~Dodonov, 
Adv. Chem. Phys. {\bf 119}, 309 (2001).

\bibitem{Dodonov2010}
V.V.~Dodonov,
arXiv:1004.3301 (2010).

\bibitem{Dalvit2010}
D.A.R.~Dalvit, P.A.~Maia Neto, and F.D.~Mazzitelli,
arXiv:1006.4790 (2010).

\bibitem{Onofrio2006}
W.-J.~Kim, J.H. Brownell, and R.~Onofrio,
Phys. Rev. Lett {\bf 96}, 200402 (2006).

\bibitem{Crocce2004}
M.~Crocce, D.A.R.~Dalvit, F.C.~Lombardo, and F.D.~Mazzitelli, 
Phys. Rev. A {\bf 70}, 033811 (2004).

\bibitem{Braggio2005}
C. Braggio, G. Bressi, G. Carugno, C. Del Noce, G. Galeazzi, A. Lombardi, A. Palmieri, G. Ruoso, and D. Zanello,  
Europhys. Lett. {\bf 70} 754 (2005).

\bibitem{Segev2007}
E.~Segev, B.~ Abdo, O.~Shtempluck, E.~Buks, and B.~Yurke,
Phys. Lett. A {\bf 370}, 202 (2007).

\bibitem{Johansson2009}
J.R.~Johansson, G.~Johansson, C.M.~Wilson, and F.~Nori,
Phys. Rev. Lett. {\bf 103}, 147003 (2009).

\bibitem{Gunter2009}
G.~Gunter, A.A.~Anappara, J.~Hees, A.~Sell, G.~Biasiol, L.~Sorba, S.~De Liberato, C.~Ciuti, A.~Tredicucci, A.~Leitenstorfer, and R.~Huber,
Nature {\bf 458}, 7235 (2009)

\bibitem{DeLiberato2009}
S.~De Liberato, D.~Gerace, I.~Carusotto, and C.~Ciuti,
Phys. Rev. A {\bf 80}, 053810 (2009).

\bibitem{You2005} 
J.Q. You and F. Nori, 
Phys. Today {\bf 58} (11), 42 (2005).

\bibitem{Wendin2006}
G. Wendin and V. Shumeiko, 
in {\it Handbook of Theoretical and Computational Nanotechnology}, ed. M. Rieth and W. Schommers (ASP, Los Angeles, 2006).

\bibitem{Clarke2008}
J. Clarke and F.K. Wilhelm,
Nature {\bf 453}, 1031 (2008).

\bibitem{Chiorescu2004}
I.~Chiorescu, P.~Bertet, K.~Semba, Y.~Nakamura, C.J.P.M.~Harmans, and J.E.~Mooij, 
Nature {\bf 431}, 159 (2004). 

\bibitem{Wallraff2004}
A.~Wallraff, D.I.~Schuster, A.~Blais, L.~Frunzio, R.-S.~Huang, J.~Majer, S.~Kumar, S.M.~Girvin, and R.J.~Schoelkopf,
Nature {\bf 431}, 162 (2004).

\bibitem{Schoelkopf2008}
R. J. Schoelkopf and S. M. Girvin, 
Nature {\bf 451}, 664 (2008).

\bibitem{Ashhab2010}
S.~Ashhab and F.~Nori,
Phys. Rev. A {\bf 81}, 042311 (2010).

\bibitem{Astafiev2007}
O.~Astafiev, K.~Inomata, A.O.~Niskanen, T.~Yamamoto, Yu.A.~Pashkin, Y.~Nakamura, and J.S.~Tsai, 
Nature {\bf 449}, 588 (2007).

\bibitem{Ashhab2008}
S.~Ashhab, J.R.~Johansson, A.M.~Zagoskin, and F.~Nori,
New J. Phys. {\bf 11}, 023030 (2009).

\bibitem{Hofheinz2008}
M.~Hofheinz, H.~Wang, M.~Ansmann, R.C.~Bialczak, E.~Lucero, M.~Neeley, A.D.~O'Connell, D.~Sank, J.~Wenner, J.M.~Martinis, and A.N.~Cleland, 
Nature {\bf 459}, 546 (2009);

\bibitem{Liu2004}
Y.X.~Liu, L.F.~Wei, and F.~Nori,
Europhys. Lett. {\bf 67}, 941–947 (2004).

\bibitem{Zhou2008}
L.~Zhou, Z.R.~Gong, Y.X.~Liu, C.P.~Sun, and F.~Nori,
Phys. Rev. Lett. {\bf 101}, 100501 (2008).

\bibitem{Abdumalikov2008}
A.A.~Abdumalikov, O.~Astafiev, Y.~Nakamura, Y.A.~Pashkin, and J.S.~Tsai,
Phys. Rev. B {\bf 78}, 180502 (2008)

\bibitem{Astafiev2010}
O.~Astafiev, A.M.~Zagoskin, A. M., A.A.~Abdumalikov, Y.A.~Pashkin, T.~Yamamoto, K.~Inomata, Y.~Nakamura, and J.S.~Tsai,
Science {\bf 327}, 840 (2010)

\bibitem{Liao2010}
J.-Q.~Liao, Z.R.~Gong, L.~Zhou, Y.X.~Liu, C.P.~Sun, and F.~Nori,
Phys. Rev. A {\bf 81}, 042304 (2010)

\bibitem{Palacois2008}
A. Palacios-Laloy, F. Nguyen, F. Mallet, P. Bertet, D. Vion, and D. Esteve,
J. Low Temp. Phys. {\bf 151}, 1034 (2008).

\bibitem{Yamamoto2008}
T.~Yamamoto, K.~Inomata, M.~Watanabe, K.~Matsuba, T.~Miyazaki, W.D.~Oliver, Y.~Nakamura, and J.S.~Tsai, 
Appl. Phys. Lett. {\bf 93}, 042510 (2008).

\bibitem{Lehnert2008}
M.A. Castellanos-Beltran, K.D. Irwin, G.C. Hilton, L.R. Vale, and K.W. Lehnert, 
Nat. Phys. {\bf 4}, 929 (2008).

\bibitem{Sandberg2008}
M.~Sandberg, C.M.~Wilson, F.~Persson, T.~Bauch, G.~Johansson, V.~Shumeiko, T.~Duty, and P.~Delsing, 
Appl. Phys. Lett. {\bf 92}, 203501 (2008).

\bibitem{Buluta2009}
I.~Buluta and F.~Nori,
Science {\bf 326}, 5949 (2009).

\bibitem{Takashima2008}
K.~Takashima, N.~Hatakenaka, S.~Kurihara, and A.~Zeilinger, 
J.~Phys.~A: Math.~Theor. 
{\bf 41} 164036 (2008).

\bibitem{Nation2009} 
P.D.~Nation, M.P.~Blencowe, A.J.~Rimberg, and E.~Buks, 
Phys.~Rev.~Lett.~{\bf 103}, 087004 (2009).

\bibitem{Crispino2008}
L.C.B.~Crispino, A.~Higuchi, and G.E.A.~Matsas,
Rev. Mod. Phys. {\bf 80}, 787 (2008).

\bibitem{Casimir1948}
H.B.G.~Casimir,
Proc.~K.~Ned.~Akad.~Wet. {\bf 51}, 793 (1948).

\bibitem{Lifshitz1956}
E.M.~Lifshitz,
Sov.~Phys.~JETP  {\bf 2}, 73 (1956).

\bibitem{Sparnaay1958}
M.J.~Sparnaay,
Physica {\bf 24}, 751 (1958).

\bibitem{vanBlokland1978}
P.H.G.M.~van Blokland and J.T.G.~Overbeek,
J. Chem. Soc., Faraday Trans., {\bf 74}, 2637 (1978).

\bibitem{Lamoreaux1997}
S.K.~Lamoreaux,
Phys. Rev. Lett. {\bf 78}, 5 (1997).

\bibitem{Mohideen1998}
U.~Mohideen and A.~Roy,
Phys. Rev. Lett. {\bf 81}, 4549 (1998).

\bibitem{Bressi2002}
G.~Bressi, G.~Carugno, R.~Onofrio, and G.~Ruoso,
Phys. Rev. Lett. {\bf 88}, 041804 (2002).

\bibitem{Milonni1994}
P.W. Milonni,
{\it The Quantum Vacuum: An Introduction to Quantum Electrodynamics}
(Academic, San Diego, 1994).

\bibitem{Lamoreaux1999}
S.K.~Lamoreaux,
Am.~J.~Phys.~{\bf 67}, 850 (1999).

\bibitem{Capasso2007}
F.~Capasso, J.N.~Munday, D.~Iannuzzi, and H.B.~Chan,
IEEE J.~Sel.~Top.~Quantum Electron.~{\bf 13}, 400 (2007).

\bibitem{Jaekel1992}
M.-T.~Jaekel and S.~Reynaud,
J.~Phys.~I {\bf 2}, 149 (1992).

\bibitem{Dodonov1990}
V.V.~Dodonov,
Phys.~Lett.~A. {\bf 149}, 225 (1990).

\bibitem{Dodonov1993}
V.V.~Dodonov, A.B.~Klimov, and D.E.~Nikonov,
J.~Math.~Phys. {\bf 34}, 7 (1993).

\bibitem{Law1994}
C.K.~Law,
Phys.~Rev.~Lett. {\bf 73}, 1931 (1994).

\bibitem{Cole1995}
C.C.~Cole and W.C.~Schieve,
Phys.~Rev.~A {\bf 52}, 4405 (1995).

\bibitem{Dalvit1998}
D.A.R.~Dalvit and F.D.~Mazzitelli,
Phys.~Rev.~A {\bf 57}, 2113 (1998)

\bibitem{Dalvit1999}
D.A.R.~Dalvit and F.D.~Mazzitelli,
Phys.~Rev.~A {\bf 59}, 3049 (1999)

\bibitem{Razavy1989}
M.~Razavy and J.~Terning,
Phys.~Rev.~D {\bf 31}, 307 (1989).

\bibitem{Law1993}
C.K.~Law,
Phys.~Rev.~A {\bf 49}, 433 (1994).

\bibitem{Law1995}
C.K.~Law,
Phys.~Rev.~A {\bf 51}, 2537 (1995).

\bibitem{Dodonov1995}
V.V.~Dodonov,
Phys.~Lett.~A {\bf 207}, 126 (1995).

\bibitem{Dodonov1996}
V.V.~Dodonov and A.B.~Klimov,
Phys.~Rev.~A {\bf 53}, 2664 (1996).

\bibitem{Crocce2001}
M.~Crocce, D.A.R.~Dalvit, and F.D.~Mazzitelli,
Phys.~Rev.~A {\bf 64}, 013808 (2001).

\bibitem{Alves2006}
D.T.~Alves, C.~Farina, and E.R.~Granhen,
Phys.~Rev.~A {\bf 73}, 063818 (2006).

\bibitem{Alves2010}
D.T.~Alves, E.R.~Granhen, H.O.~Silva, and M.G.~Lima,
Phys.~Rev.~D {\bf 81}, 025016 (2010).

\bibitem{Meplan1995}
O.~Meplan and C.~Gignoux,
Phys.~Rev.~Lett {\bf 76}, 408 (1995).

\bibitem{Lambrecht1996}
A.~Lambrecht, M.-T.~Jaekel, and S.~Reynaud,
Phys.~Rev.Lett. {\bf 77}, 615 (1996).

\bibitem{Lambrecht1998}
A.~Lambrecht, M.-T.~Jaekel, and S.~Reynaud,
Euro.~Phys.~J.~D {\bf 3}, 95 (1998).

\bibitem{Neto1996}
P.A.~Maia Neto and L.A.S.~Machado,
Phys.~Rev.~A {\bf 54}, 3420 (1996).

\bibitem{Yurke1984}
B.~Yurke and J.S.~Denker, 
Phys. Rev. A {\bf 29}, 1419 (1984).

\bibitem{Devoret1995}
M.~Devoret, 
p.~351-386, (Les Houches LXIII, 1995) (Amsterdam: Elsevier).

\bibitem{Wallquist2006}
M.~Wallquist, V.S.~Shumeiko, and G.~Wendin, 
Phys. Rev. B {\bf 74}, 224506 (2006).

\bibitem{Likharev1986}
K.K.~Likharev,
{\it Dynamics of Josephson junctions and circuits}
(Gordon, Amsterdam, 1986).

\bibitem{Loudon1987}
R.~Loudon and P.L.~Knight,
J.~Mod.~Optics {\bf 34}, 709 (1987).

\bibitem{Caves1985}
C.M.~Caves and B.L.~Schumaker,
Phys. Rev. A {\bf 31}, 3068 (1985).

\bibitem{Zagoskin2008}
A.M.~Zagoskin, E.~Il'ichev, M.W.~McCutcheon, J.F.~Young, and F.~Nori,
Phys. Rev. Lett. {\bf 101}, 253602 (2008).

\bibitem{Walls1994}
D.F.~Walls and G.J.~Milburn, 
{\it Quantum Optics}
(Springer, Berlin, 1994).

\bibitem{Glauber1963}
R.J.~Glauber, Phys. Rev. {\bf 130}, 2529 (1963);
R.J.~Glauber, Phys. Rev. {\bf 131}, 2766 (1963).

\bibitem{Bozyigit2010}
D.~Bozyigit, C.~Lang, L.~Steffen, J.M.~Fink, M.~Baur, R.~Bianchetti, P.J.~Leek, S.~Filipp, M.P.~da Silva, A.~Blais, and A.~Wallraff,
arXiv:1002.3738 (2010).

\bibitem{Caves1982}
C.M.~Caves, Phys.~Rev.~D {\bf 26}, 1817 (1982).

\bibitem{Clerk2010}
A.A.~Clerk, M.H.~Devoret, S.M.~Girvin, F.~Marquardt, and R.J.~Schoelkopf,
Rev.~Mod.~Phys. {\bf 82}, 1155 (2010).

\bibitem{Dezael2009}
F.X.~Dezael and A.~Lambrecht,
0912.2853v1 (2009).

\bibitem{Wilson2010}
C.M.~Wilson, T.~Duty, M.~Sandberg, F.~Persson, V.~Shumeiko, and P.~Delsing,
arXiv:1006.2540 (2010).

\end{thebibliography}
\end{document}